\documentclass[aps,prl,twocolumn,showpacs,citeautoscript,superscriptaddress]{revtex4-1}

\usepackage{graphicx}  
\usepackage{dcolumn}   
\usepackage{bm}        
\usepackage{amssymb}   
\usepackage[none]{hyphenat}  
\usepackage{nicefrac}  
\usepackage{amsmath}   
\usepackage{footmisc}  
\usepackage{latexsym}  
\usepackage{romannum}  
\usepackage{hyperref}  
\usepackage{multirow}  
\usepackage{array}

\usepackage[table]{xcolor}
\usepackage{color,colortbl}
\definecolor{yellow}{rgb}{1,1,0.5}
\usepackage{tabulary}

\renewcommand{\arraystretch}{1.5}

\newcommand{\Ev}{\mathbf{E}}
\newcommand{\Hv}{\mathbf{H}}
\newcommand{\Bv}{\mathbf{B}}
\newcommand{\Dv}{\mathbf{D}}
\newcommand{\ev}{\hat{\mathbf{e}}}
\newcommand{\kv}{\mathbf{k}}
\newcommand{\Rv}{\mathbf{R}}

\newcommand{\kb}{\overline{\overline{\mathbf{k}}}}
\newcommand{\epsb}{\overline{\overline{\varepsilon}}}
\newcommand{\mub}{\overline{\overline{\mu}}}
\newcommand{\xib}{\overline{\overline{\xi}}}
\newcommand{\zetab}{\overline{\overline{\zeta}}}

\renewcommand{\Im}{\mathrm{Im}}

\newcommand{\eps}{\varepsilon}

\begin{document}

\title{Universal spin-resolved thermal radiation laws for
  nonreciprocal bianisotropic media}

\author{Chinmay Khandekar} \email{ckhandek@purdue.edu}
\affiliation{Birck Nanotechnology Center, School of Electrical and
  Computer Engineering, College of Engineering, Purdue University,
  West Lafayette, Indiana 47907, USA}

\author{Farhad Khosravi} \affiliation{Department of Electrical and
  Computer Engineering, University of Alberta, Edmonton, Alberta
  T6G1H9, Canada}

\author{Zhou Li} 
\affiliation{Birck Nanotechnology Center, School of Electrical and
  Computer Engineering, College of Engineering, Purdue University,
  West Lafayette, Indiana 47907, USA}
    
\author{Zubin Jacob} \email{zjacob@purdue.edu}
\affiliation{Birck Nanotechnology Center, School of Electrical and
  Computer Engineering, College of Engineering, Purdue University,
  West Lafayette, Indiana 47907, USA}

\date{\today}

\begin{abstract}
  A chiral absorber of light can emit spin-polarized (circularly
  polarized) thermal radiation based on Kirchhoff's law which equates
  spin-resolved emissivity with spin-resolved absorptivity for
  reciprocal media at thermal equilibrium. No such law is known for
  nonreciprocal media. In this work, we discover three spin-resolved
  Kirchhoff's laws of thermal radiation applicable for both reciprocal
  and nonreciprocal planar media. In particular, these laws are
  applicable to multi-layered or composite slabs of generic
  bianisotropic material classes which include (uniaxial or biaxial)
  birefringent crystals, (gyrotropic) Weyl semimetals, magnetized
  semiconductors, plasmas, ferromagnets and ferrites,
  (magnetoelectric) topological insulators, metamaterials and
  multiferroic media. We also propose an experiment to verify these
  laws using a single system of doped Indium Antimonide (InSb) thin
  film in an external magnetic field. Furthermore, we reveal a
  surprising result that the planar slabs of all these material
  classes can emit partially circularly polarized thermal light
  without requiring any surface patterning, and identify planar
  configurations which can experience nontrivial thermal
  optomechanical forces and torques upon thermal emission into the
  external environment at lower temperature (nonequilibrium). Our work
  also provides a new fundamental insight of detailed balance of
  angular momentum (in addition to energy) of equilibrium thermal
  radiation, and paves the way for practical functionalities based on
  thermal radiation using nonreciprocal bianisotropic materials.
\end{abstract}

 \maketitle

\section{Introduction}

At the foundation of the field of thermal radiation lies Kirchhoff's
law which relates emissivity with absorptivity. One formulation of
Kirchhoff's law in the context of circularly polarized light states
that emissivity ($\eta$) is equal to absorptivity ($\alpha$) for both
left circular polarization (LCP) and right circular polarization (RCP)
states. Its mathematical form is:
\vspace{-0.2cm}
\begin{align} 
\eta_{(+,-)}(\omega,\mathbf{n}) =
\alpha_{(+,-)}(\omega,\mathbf{n}) \nonumber
\end{align} 
where $(+)$ denotes RCP, $(-)$ denotes LCP, $\omega$ is the frequency
and $\mathbf{n}$ denotes the direction. However, this spin-resolved
Kirchhoff's law is valid only for reciprocal media. Many works have
used it to design reciprocal chiral
absorbers~\cite{wu2014spectrally,yin2013interpreting,dyakov2018magnetic}
which can emit partially spin-polarized thermal radiation and few
works have demonstrated it in
experiments~\cite{shitrit2013spin,wadsworth2011broadband}. This
conventional law is not applicable for nonreciprocal media with broken
time reversal symmetry such as semiconductors in external magnetic
fields. Naturally, the question then arises whether new forms of
Kirchhoff's laws exist~\cite{miller2017universal}. In this work, we
provide the spin-resolved Kirchhoff's laws which are applicable for
nonreciprocal media.

Thermal radiation from nonreciprocal media is an emerging research
area~\cite{zhu2014near,moncada2015magnetic,latella2017giant,
  ben2016photon,herz2019green}. There are no experiments till date and
computational tools for analyzing nonreciprocal thermal radiation in
arbitrary geometries are still under
development~\cite{ekeroth2017thermal,zhu2018theory}. While the recent
interesting works primarily investigate the heat flux, we introduce a
different perspective of spin-resolved or spin-polarized radiative
heat flux, which is partially motivated by the well-known concept of
electronic spin currents in condensed matter
physics~\cite{RevModPhys.76.323}. In our recent
work~\cite{khandekar2019spin}, we comprehensively analyze the spin of
thermal radiation in the \emph{near-field} of generic nonreciprocal
media. We show interesting effects like persistent radiative heat
current and persistent photon spin that can exist in the near-field
without any temperature difference and explain their connection with
the spin-momentum locking of evanescent photonic
states~\cite{van2016universal}. Here, we provide a similar
comprehensive analysis in the \emph{far-field} of nonreciprocal media.

We derive the spin-resolved emissivities and absorptivities for planar
media using the radiometry principles and based on fluctuational
electrodynamic calculations~\cite{khandekar2019spin}. Our derivation
is applicable for both reciprocal and nonreciprocal media because it
does not require the concept of electromagnetic reciprocity. We
further validate this derivation by proving thermodynamic consistency
based on the detailed balance of linear and angular momentum transfer,
via thermal radiation between the planar slab and the environment at
thermal equilibrium. The spin-resolved emissivities and absorptivities
ensure that there is no nonzero force (linear momentum balance) or
torque (angular momentum balance) on the planar slab when it is at
thermal equilibrium with the environment.

The principle of detailed balance of energy plays a pivotal role in
the derivation of many heat-transfer laws. Here, we introduce a new
concept of detailed balance of angular momentum of radiative heat at
thermal equilibrium. Since this concept is generalizable to electrons,
phonons, photons in various systems, it represents a paradigm shift in
the analysis of heat transfer at thermal equilibrium. In this work, it
is crucial for validating the spin-resolved emissivities for
nonreciprocal media which have not been derived previously.

We then analyze the spin-resolved thermal emission and absorption for
planar media of all time-invariant bianisotropic material classes. The
reciprocal media include uni/biaxial anisotropic
materials~\cite{yariv1984optical} and reciprocal magnetoelectric
(chiral) media~\cite{wang2016optical,asadchy2018bianisotropic}, apart
from commonly considered isotropic dielectric and metallic
materials. The nonreciprocal materials include gyroelectric media like
metals and semiconductors in magnetic
field~\cite{ishimaru2017electromagnetic,
  armelles2013magnetoplasmonics,electron2019sengupta} and Weyl
semimetals~\cite{kotov2018giant}, gyromagnetic media like ferromagnets
and ferrites~\cite{rodrigue1988generation}, and nonreciprocal
magnetoelectric media like
multiferroics~\cite{pyatakov2012magnetoelectric}, topological
insulators~\cite{laforge2010optical} and magnetoelectric
heterostructures~\cite{hu2017understanding}. The universal
spin-resolved analysis of these material classes leads us to discover
three spin-resolved Kirchhoff's laws ({\bf SKL}s) which provide useful
relations between the spin-resolved emissivities and
absorptivities. While the first law for the reciprocal media is
intuitively expected, the other two laws for the nonreciprocal media
are new. They demonstrate a subtle balance of spin-resolved emission
and absorption of thermal radiation, which maintains thermal
equilibrium of the non-reciprocal planar media with the
surrounding. We emphasize that these laws are applicable for composite
or multi-layered configurations of these materials for all
frequencies, emission directions, material and geometry parameters. We
further propose an experiment which can validate these laws
conveniently using a single material system of a doped InSb thin film
in the presence of magnetic field.

Apart from the Kirchhoff's laws derived at thermal equilibrium, we
also reveal other interesting spin-resolved thermal radiation features
observed when the external environment is at lower temperature
(nonequilibrium condition). First, we point out a striking result that
the planar geometries of all material classes mentioned above except
the conventional isotropic dielectric/metallic materials can emit
partially spin-polarized thermal radiation in suitable
directions. This is important from the perspective of generating
circularly polarized light because most of these materials do not
require any surface patterning which otherwise causes additional
design complexity in previous
experiments~\cite{shitrit2013spin,wadsworth2011broadband}. Second, we
show that the planar slabs can experience nonequilibrium force and
torque due to the loss of linear momentum and angular momentum
respectively via thermal emission. We discuss their directionalities
for all material classes so that both parallel and perpendicular
components of the thermal-nonequilibrium optomechanical forces and
torques can be engineered using multi-layered planar slabs.

All these findings and theory details will be useful for engineering
directional radiative heat
transfer~\cite{khandekar2019spin,zhu2014near} for optimized energy
harvesting~\cite{green2012time}, designing novel spin-polarized
LEDs~\cite{khandekar2019circularly}, tailoring thermal optomechanical
forces and torques~\cite{maghrebi2019fluctuation,reid2017photon}, and
classifying or identifying materials based on infrared
polarimetry. Fundamentally, our work simultaneously advances many
research topics like thermal radiation, photon spin and angular
momentum, electromagnetic nonreciprocity, bianisotropic media and
metamaterials, optical characterization of materials etcetera.
  
\section{Results}
  
\subsection{Spin-resolved Kirchhoff's laws}
\begin{figure*}[t!]
  \centering\includegraphics[width=\linewidth]{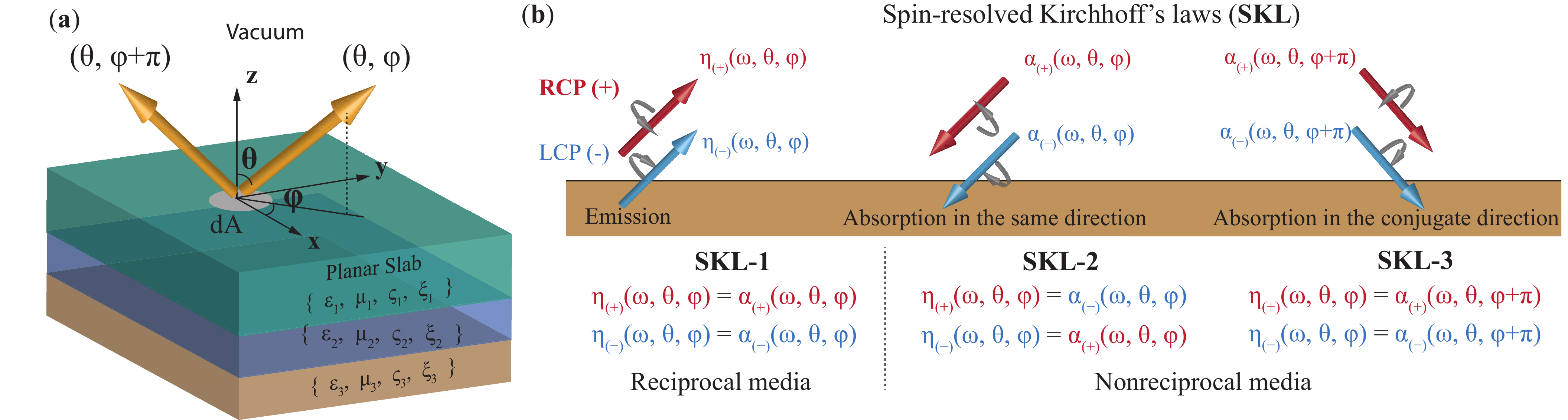}
  \caption{(a) The figure shows the conventions used to describe the
    thermal radiation from the planar slab in a specific emission
    direction denoted by the angles ($\theta,\phi$). Figure (b)
    summarizes the three spin-resolved Kirchhoff's laws that relate
    spin-resolved emissivities in ($\theta,\phi$) direction with
    spin-resolved absorptivities either along the same direction
    ($\theta,\phi$) or along the conjugate direction denoted by the
    angles ($\theta,\phi+\pi$). The directions are consistently
    characterized based on the emission direction and not the
    absorption or wavevector directions. These laws are applicable for
    reciprocal and nonreciprocal materials presented in the
    table~\ref{table1} and their multilayered or composite
    configurations as described in the main text.}
  \vspace{0.1in}
  \label{fig1}
\end{figure*}

We consider an extended finite-thickness multilayered planar slab much
larger than the thermal wavelength at thermal equilibrium with the
blackbody radiation at temperature $T$. As shown in fig.\ref{fig1}(a),
the power emitted per unit surface area ($dA$) of the body at an angle
$\theta$ to the surface normal within the solid angle $d\Omega$ per
unit frequency interval ($d\omega$) is given by,
\begin{align}
P_{\text{rad}}(\omega,\theta,\phi,\ev) =
\eta(\omega,\theta,\phi,\ev)\frac{I_b(\omega,T)}{2} \cos\theta d\omega
d\Omega dA
\label{emission}
\end{align}
where $\eta$ is the dimensionless emissivity dependent on frequency
($\omega$), emission direction ($\theta,\phi$) in spherical
coordinates and orthogonal polarization states ($\ev$). The angle
$\phi\in [0,2\pi]$ while the angle $\theta\in [0,\pi]$ such that
$\theta \in [0,\pi/2)$ denotes the top hemisphere and $\theta \in
  [\pi/2,\pi]$ denotes the bottom hemisphere. For a semi-infinite
  half-space, $\theta \in [0,\pi/2]$.  The blackbody radiance at
  temperature $T$ given by
  $I_b(\omega,T)=\omega^2\Theta(\omega,T)/(4\pi^3 c^2)$ with
  $\Theta(\omega,T)=\hbar\omega/[\text{exp}(\hbar\omega/k_B T)-1]$
  being the Planck's function, is divided by two to account for two
  polarization states separately. Similarly, the power absorbed per
  unit surface area $dA$ due to the blackbody radiance incident at an
  angle $\theta$ to the surface normal within a solid angle $d\Omega$
  per unit frequency interval $d\omega$ is,
\begin{align}
P_{\text{abs}}(\omega,\theta,\phi,\ev) =
\alpha(\omega,\theta,\phi,\ev)\frac{I_b(\omega,T)}{2} \cos\theta
d\omega d\Omega dA
\label{absorption}
\end{align}
We focus on polarization-dependent properties of thermal
emission. Instead of the usual $s,p$-polarization basis, we consider
RCP ($\ev_+$) and LCP ($\ev_-$) polarization states. The circular
polarization basis states in the basis of s,p polarization states
(${\ev_s, \ev_p}$) are $\ev_{\pm} = (\ev_s \pm i\ev_p)/\sqrt{2}$.

\begin{table*}
  \centering \setlength{\tabcolsep}{0.15cm}
  {\renewcommand{\arraystretch}{0.9}
  \begin{tabular}{| c | p{2.3cm} | p{5cm} | p{6cm} | c |}
    \hline No. & Material Class & Description & Examples & Kirchhoff's
    law \\ \hline \hline $1$ & Reciprocal isotropic & $\epsb, \mub$
    are scalars, $\xib=\zetab=0$ & Common dielectric and metallic
    materials & {\bf SKL}-1 \\ \hline $2$ & Reciprocal anisotropic &
    Diagonal $\epsb$ with unequal entries, scalar $\mub$,
    $\xib=\zetab=0$. & Uniaxial and biaxial
    crystals~\cite{yariv1984optical} & {\bf SKL}-1 \\ \hline $3$ &
    Nonreciprocal gyroelectric & $\epsb \neq \epsb^{T}$, scalar
    $\mub$, $\xib=\zetab=0$. & Weyl semimetals~\cite{kotov2018giant},
    metals and semiconductors in magnetic
    field~\cite{ishimaru2017electromagnetic,
      armelles2013magnetoplasmonics,electron2019sengupta} & {\bf
      SKL}-2,3 \\ \hline $4$ & Nonreciprocal gyromagnetic & $\mub \neq
    \mub^{T}$, scalar $\epsb$, $\xib=\zetab=0$. & Ferromagnets,
    ferrites~\cite{rodrigue1988generation} & {\bf SKL}-2,3 \\ \hline
    $5$ & Reciprocal magneto-electric & scalar $\epsb$ and $\mub$,
    nonzero $\xib = -\zetab^T$. & Chiral media,
    metamaterials~\cite{wang2016optical,asadchy2018bianisotropic} &
    {\bf SKL}-1 \\ \hline $6$ & Nonreciprocal magneto-electric &
    scalar $\epsb$ and $\mub$, nonzero $\xib = \zetab^T$. &
    Topological insulator~\cite{laforge2010optical}, Multi-ferroic
    media~\cite{pyatakov2012magnetoelectric} and
    heterostructures~\cite{hu2017understanding} & {\bf SKL}-2,3
    \\ \hline
  \end{tabular}}
\caption{Reciprocal and nonreciprocal material classes}
\label{table1}
\end{table*}

We derive (as explained in detail in the discussion section) and
analyze the spin-resolved emissivities
$\eta_{(\pm)}=\eta(\omega,\theta,\phi,\ev_{\pm})$ and absorptivities
$\alpha_{(\pm)}= \alpha(\omega,\theta,\phi,\ev_{\pm})$ defined above
for planar configurations of several material classes. As a superset
of materials, we consider a linear, time-invariant generic
bianisotropic medium whose optical properties are described using the
following constitutive relations in the frequency domain assuming
local material response:
\begin{align}
  \Dv = \epsb\eps_0 \Ev + \xib\frac{1}{c}\Hv, \hspace{10pt} \Bv =
  \zetab\frac{1}{c}\Ev + \mub\mu_0\Hv
  \label{consti}
\end{align}
where $\epsb,\mub$ are dimensionless permittivity and permeability
tensors and $\xib,\zetab$ are magneto-electric coupling tensors. Such
a material is reciprocal if these parameters satisfy the conditions $
\epsb=\epsb^{T}, \mub=\mub^{T}, \xib=-\zetab^{T}$ where $(\cdots)^{T}$
denotes the matrix transpose. It is nonreciprocal if any one of these
conditions is violated. We consider a list of several material classes
as shown in table~\ref{table1}, which includes isotropic, uni/biaxial
anisotropic, gyroelectric, gyromagnetic and magneto-electric
materials. We discover that for multilayered or composite planar slabs
of most of these material classes, the spin-resolved emissivities
($\eta_{(\pm)}$) and absorptivities ($\alpha_{(\pm)}$) satisfy
specific relations described below and summarized in
fig~\ref{fig1}. They are called as spin-resolved Kirchhoff's laws
({\bf SKL}s).

{\bf SKL-1}: For multilayered or composite planar slabs of all
reciprocal media, the spin-resolved emissivity in ($\theta,\phi$)
direction is equal to the spin-resolved absorptivity in \emph{the same
  direction for each spin (polarization) state}.
\vspace{-1.5pt}
\begin{align}
  \eta_{(+)}(\omega,\theta,\phi) = \alpha_{(+)}(\omega,\theta,\phi)
  \nonumber \\
  \eta_{(-)}(\omega,\theta,\phi) = \alpha_{(-)}(\omega,\theta,\phi)
\label{KL1}
\end{align}

{\bf SKL-2}: For multilayered or composite planar slabs of
nonreciprocal media \emph{that preserve the rotational symmetry in the
  plane of the surface} such as gyrotropic media with gyrotropy axis
perpendicular to the surface and isotropic magneto-electric (Tellegen)
media, the spin-resolved emissivity in ($\theta,\phi$) direction is
equal to the spin-resolved absorptivity in \emph{the same direction
  but of opposite spin state}.
\vspace{-1.5pt}
\begin{align}
  \eta_{(+)}(\omega,\theta,\phi) = \alpha_{(-)}(\omega,\theta,\phi)
  \nonumber \\ \eta_{(-)}(\omega,\theta,\phi) =
  \alpha_{(+)}(\omega,\theta,\phi)
\label{KL2}
\end{align}

{\bf SKL-3}: For multi-layered or composite planar slabs of
nonreciprocal media that \emph{do not preserve the rotational
  symmetry} in the plane of the surface such as gyrotropic media with
gyrotropy axis parallel to the surface and anisotropic nonreciprocal
magneto-electric media that cause cross coupling between perpendicular
and parallel components of $\Ev-\Hv$ fields ($E_x-H_z$, $E_z-H_x$,
$E_y-H_z$ $E_z-H_y$), the spin-resolved emissivity in the direction
($\theta,\phi$) is equal to the spin-resolved absorptivity for
\emph{the conjugate direction} ($\theta,\phi+\pi$) \emph{for each spin
  state}.
\vspace{-1.5pt}
\begin{align}
\eta_{(+)}(\omega,\theta,\phi) = \alpha_{(+)}(\omega,\theta,\phi+\pi)
\nonumber \\
\eta_{(-)}(\omega,\theta,\phi) = \alpha_{(-)}(\omega,\theta,\phi+\pi)
\label{KL3}
\end{align}

We emphasize that a multilayered or a composite planar slab which
combines different material types following the same {\bf SKL} also
follows that same {\bf SKL}. For instance, a planar slab which
combines materials exhibiting gyrotropic and magnetoelectric response
simultaneously, will follow {\bf SKL}-2 or 3 depending on the
gyrotropy axis and the isotropic or anisotropic nature of the
magnetoelectric response as described above. For conventional
isotropic dielectric/metallic materials, all three laws hold since the
spin-resolved emissivities and absorptivities are equal
[$\eta_{(+)}=\eta_{(-)}=\alpha_{(+)}=\alpha_{(-)}$] in all relevant
directions. It follows that a multilayered planar slab which contains
layers of such trivial materials (satisfying all three {\bf SKL}s)
along with more exotic bianisotropic media satisfying a specific {\bf
  SKL} can be readily described using that specific {\bf
  SKL}. However, the laws are not applicable when the nontrivial
bianisotropic media following distinct {\bf SKL}s are combined. One
example can be a planar slab containing layers of uniaxial anisotropic
media ({\bf SKL}-1) and gyrotropic media ({\bf SKL}-2 or
3). Interestingly, we also find that a nonreciprocal magnetoelectric
medium that leads to coupling between the parallel components of
$\Ev-\Hv$ fields ($E_x-H_y$ or $E_y-H_x$) does not satisfy any one of
these three {\bf SKL}s. This special material is well-known in the
literature for causing Fiegel effect and related
phenomena~\cite{feigel2004quantum,van2006momentum,
  ole2007extraction}. Its uniqueness is related to the momentum
asymmetry of light propagation inside this medium and it will be
analyzed in more detail in our future work. Despite the few limiting
cases, we note that {\bf SKL}s are useful for thermal-radiation
engineering using multi-layered or composite configurations of most
bianisotropic materials, for all frequencies, emission directions,
material and geometry parameters. In the following, we propose an
experiment to verify these laws using a single material system.

\subsection{Experimental proposal}

We consider a doped Indium Antimonide (InSb) slab of thickness
$t=1\mu$m. and doping concentration $n=10^{17}$cm$^{-3}$ (available at
vendors like MTI Corp.) on top of a glass substrate of constant
permittivity $\varepsilon_s=2.25$. We use the experimentally
well-characterized Drude-Lorentz oscillator
model~\cite{kushwaha2001plasmons,palik1976coupled,chochol2016magneto}
to calculate the permittivity of InSb with or without applied magnetic
field~\cite{khandekar2019spin}. In the absence of magnetic field, InSb
slab has isotropic permittivity ($\epsb$) and permeability ($\mub$),
and $\xib,\zetab=0$. Therefore, it is reciprocal in nature and follows
{\bf SKL}-1. In the presence of magnetic field, $\epsb$ has nonzero
off-diagonal entries and InSb slab acts as a gyroelectric medium with
the gyrotropy axis parallel to the applied field. Consequently, when
the magnetic field is perpendicular or parallel to the surface,
thermal radiation from InSb slab will follow either {\bf SKL}-2 or
{\bf SKL}-3 respectively. In an experiment, the measurement of
spin-resolved emissivities (Eq.\ref{eta}) and absorptivities
(Eq.\ref{alpha}) can be performed using a specific combination of
polarizers and quarter-wave-plate optical
components~\cite{wadsworth2011broadband} in suitable directions of
emission. Here, we calculate them as a function of frequency.

\begin{figure}[t!]
  \centering\includegraphics[width=0.95\linewidth]{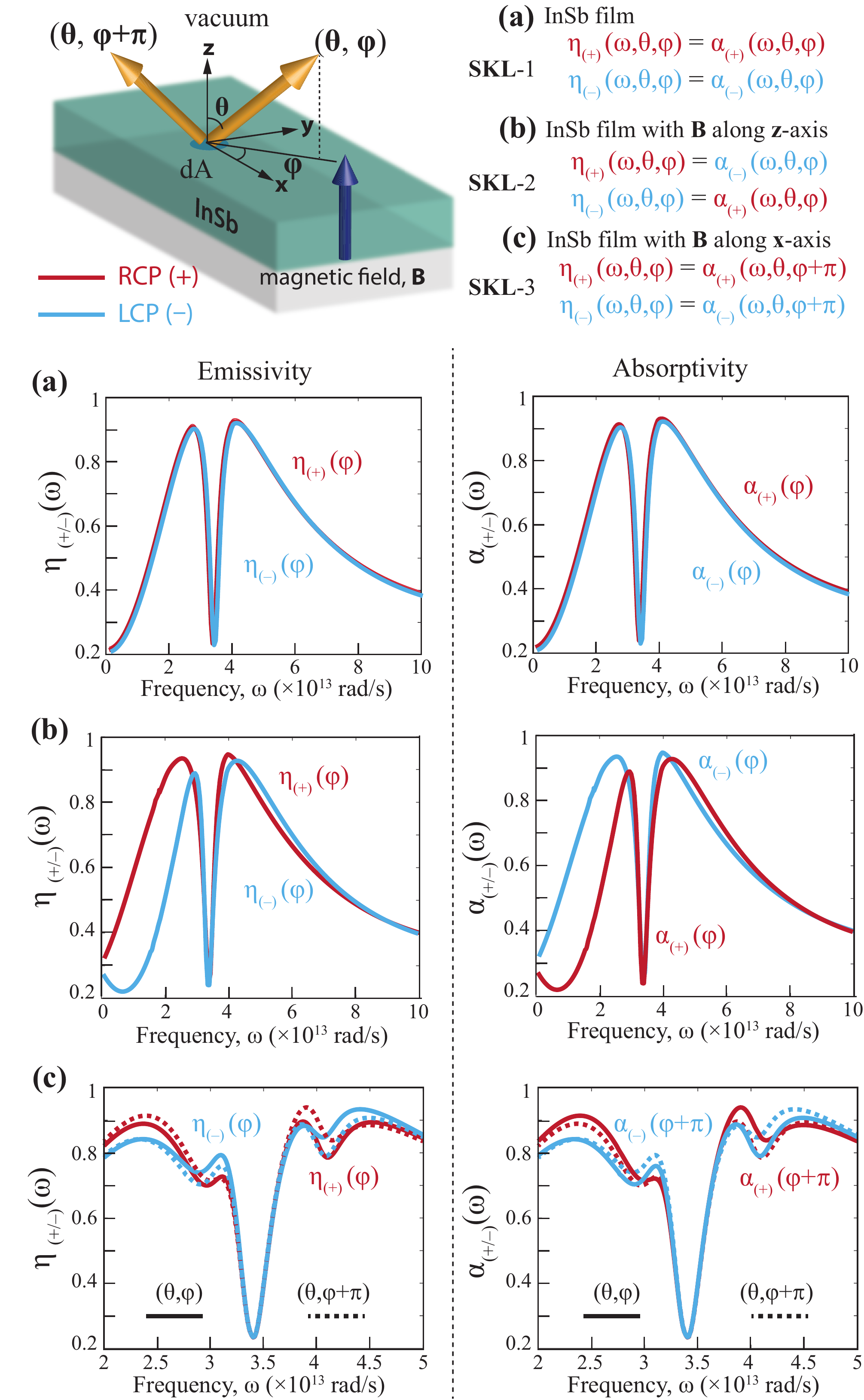}
  \caption{The top schematic summarizes the three spin-resolved
    Kirchhoff's laws for a doped InSb slab with or without externally
    applied magnetic field. Figures (a,b,c) respectively demonstrate
    these laws. For brevity, we fix the direction as
    ($\theta=\pi/4,\phi=\pi/4$). The first and second spin-resolved
    Kirchhoff's laws demonstrated in (a,b) respectively relate
    emissivities with absorptivities in the same direction. The third
    law demonstrated in (c) relates emissivity in a given direction
    with absorptivity in the conjugate direction
    ($\theta=\pi/4,\phi=\pi/4+\pi$).}
  \label{fig2}
\end{figure} 

Figure~\ref{fig2} summarizes the three spin-resolved Kirchhoff's laws
for the InSb slab (top schematic) and figures (a,b,c) demonstrate the
calculated spectra of emissivities (left figures) and absorptivities
(right figures). For brevity, we focus on the direction
($\theta=\pi/4,\phi=\pi/4$). As shown in Fig.~\ref{fig2}(a), in the
absence of magnetic field, the spin-resolved emissivities and
absorptivities are equal for both spin states, RCP (red) and LCP
(blue). The plots lie on top of each other. Figure~\ref{fig2}(b)
demonstrates {\bf SKL}-2, where a magnetic field of strength $1$T is
applied perpendicular to the slab. Evidently,
$\eta_{(+)}=\alpha_{(-)}$ and $\eta_{(-)}=\alpha_{(+)}$, which
verifies {\bf SKL}-2. Since $\eta_{(+)}\neq \eta_{(-)}$ in the given
emission direction, the thermal emission from the slab will be
partially spin-polarized. In Fig.~\ref{fig2}(c), we consider a
magnetic field of strength $2$T applied along the $\bm{x}$-axis of the
geometry. This example demonstrates that spin-resolved emissivities in
($\theta,\phi$) direction are equal to spin-resolved absorptivities
for the same spin state but in the conjugate ($\theta,\phi+\pi$)
direction. Thus, {\bf SKL}-3 can be verified.

\subsection{Universal perspective of spin-resolved thermal emission}
\label{spin-emission}

In the context of spin-resolved thermal radiation, it is illuminating
to answer two fundamental questions. First, what materials can emit
spin-polarized or circularly polarized thermal radiation? And second,
what materials can experience nontrivial torque and force upon thermal
emission of spin-polarized photons? We answer both these questions
with a universal perspective of all material classes listed in the
table~\ref{table1}. Note that we consider nonequilibrium situation
unlike equilibrium thermal radiation for derivation of {\bf SKL}s.  We
focus on thermal emission from the planar slab at temperature $T_0$
into the surrounding vacuum at $T=0$K. The strength of the circular
polarization of thermal emission is characterized using the well-known
experimentally accessible Stokes $S_3$ parameter defined as:
\begin{align}
S_3(\omega,\theta,\phi)=\frac{\eta_{(+)}(\omega,\theta,\phi)-
  \eta_{(-)}(\omega,\theta,\phi)}{\eta_{(+)}(\omega,\theta,\phi)+
  \eta_{(-)}(\omega,\theta,\phi)}
\label{stokes}
\end{align}
$S_3=+1$ denotes pure RCP light while $S_3=-1$ denotes pure LCP light
in the given emission direction. When the planar slab emits thermal
radiation into the environment at lower temperature, it loses linear
and angular momentum carried away by the photons, and consequently
experiences force and torque respectively. We compute the spectral
force per unit area ($\mathbf{F}_s$ given by Eq.~\ref{Fsa}) and the
spectral torque per unit area ($\bm{\tau}_s$ given by Eq.~\ref{Tsa})
experienced by the planar slab due to the emission of photons of
frequency $\omega$ by integrating over all emission directions. The
derivation of these quantities is provided in section~\ref{analytics}.

\begin{figure*}[t!]
  \centering\includegraphics[width=0.95\linewidth]{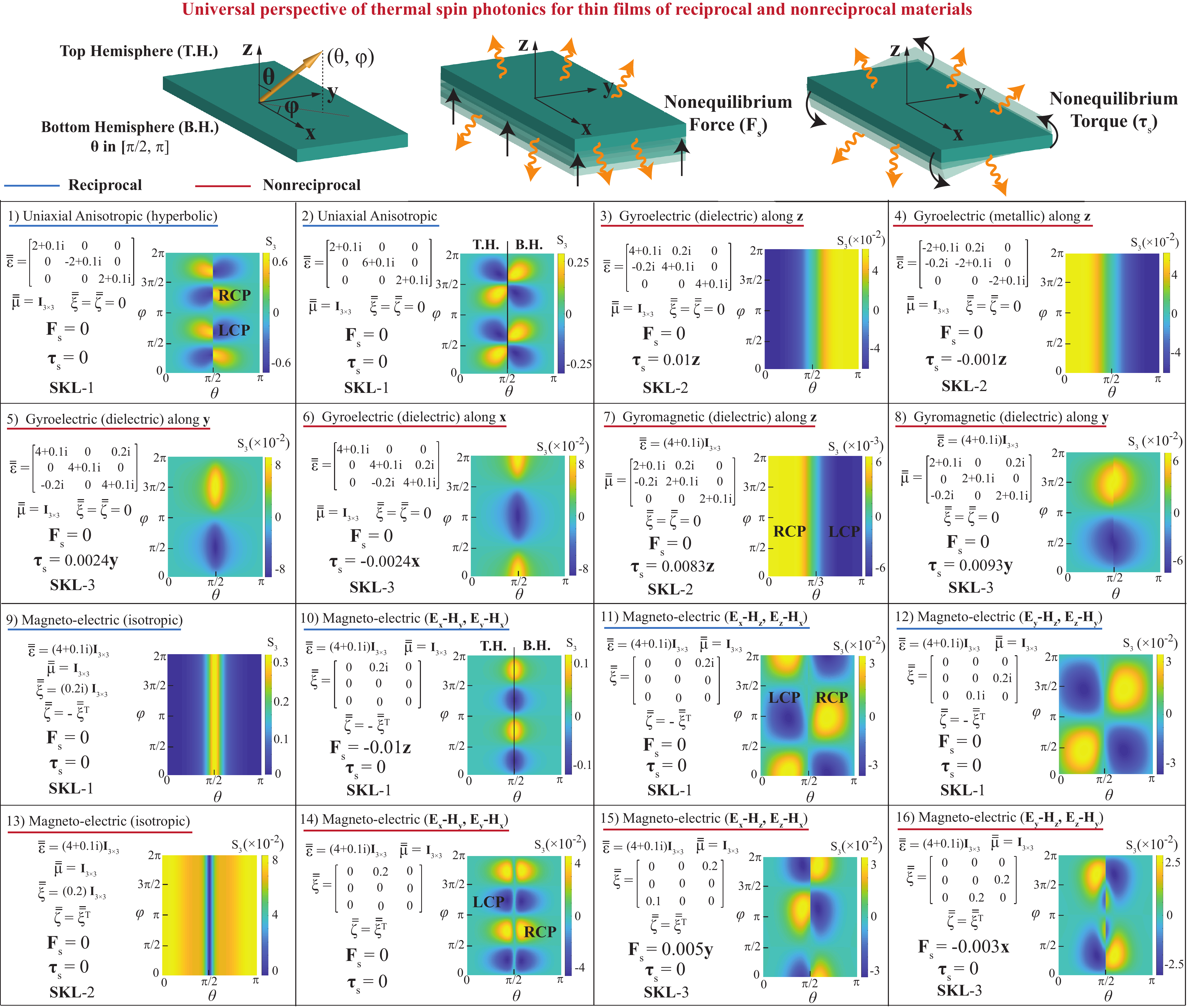}
  \caption{This figure analyzes circularly polarized thermal emission
    and associated nonequilibrium force and torque for several
    material classes using representative examples. We consider
    thermal emission at frequency $\omega$ from a planar slab of
    thickness $d=0.5c/\omega$ at temperature $T_0$ into the
    surrounding vacuum on both sides at temperature $T=0$K. The
    contour plot in each representative example demonstrates the
    Stokes $S_3$ parameter (Eq.\ref{stokes}) as a function of
    $(\theta,\phi)$ which are given in the top left inset. The
    calculated spectral force per unit area $\bm{F}_s/(I_{b,T_0}/2c)$
    and spectral torque per unit area
    $\bm{\tau}_s/(I_{b,T_0}/2\omega)$ arising from the thermal
    emission from the slab are provided for given material
    parameters. The applicable spin-resolved Kirchhoff's law ({\bf
      SKL}) is also noted for each example.}
  \label{fig3}
\end{figure*}

Figure~\ref{fig3} summarizes all materials that can emit partially
spin-polarized thermal radiation in suitable directions. For
illustration, we consider a planar slab of a material characterized by
$\epsb,\mub,\xib,\zetab$ at a temperature $T_0$ emitting into the
surrounding vacuum at $T=0$K. Similar to our previous
work~\cite{khandekar2019spin}, we provide a universal perspective by
analyzing the representative examples of several material classes
using reasonable values of the material parameters evaluated at a
frequency $\omega$ and importantly, satisfying the thermodynamic
passivity constraint~\cite{silveirinha2010comment}. While the
constitutive relations in Eq.~\ref{consti} can, in principle, be used
to describe optically active gain media, our radiometry analysis is
applicable for passive media based on the detailed balance at thermal
equilibrium. Hence, passivity of the medium must be
ensured~\cite{khandekar2019spin}. We assume the thickness of the
planar slab to be $d=0.5 c/\omega$. The contour plot for each
representative example demonstrates the variation of the Stokes $S_3$
parameter as a function of the emission direction characterized by
$(\theta,\phi)$ in spherical coordinates. The calculated spectral
force and torque per unit area provided for each example reveal the
zero or nonzero values and the directions of these quantities for all
materials belonging to that particular class. While there are many new
interesting results in this figure, a detailed discussion involving
frequency-dependent response of real materials of each type is beyond
the scope of the present work. Nonetheless, this universal perspective
is useful because we can provide a signature of new interesting
possibilities which can then be explored in more detail in the future.

{\bf Circularly polarized thermal emission}: First, we note a
surprising result that there are many materials whose planar slabs can
emit partially spin-polarized thermal radiation. This is evident from
the nonzero values of $S_3$ parameters demonstrated using contour
plots in Fig.\ref{fig3} for several material classes. This includes
all nonreciprocal planar media and also reciprocal media like
uni/bi-axial anisotropic materials and reciprocal magnetoelectric
(chiral) media. For uni-axial anisotropic media, we assume anisotropy
to be in the plane of the slab surface ($\bm{xy}$-plane). Note that
all materials belonging to these material classes except artificial
metamaterials do not require any surface patterning or geometric
chirality which is essential for observing circularly polarized
thermal radiation in previous
experiments~\cite{shitrit2013spin,wadsworth2011broadband}. Another
result is the subtle difference between spin-resolved emission from
gyrotropic media and that from nonreciprocal magnetoelectric
media. Particularly, gyrotropic media with gyrotropy axis
perpendicular to surface (examples 3,4,7) and nonreciprocal isotropic
magnetoelectric (Tellegen) media (example 13) both satisfy {\bf SKL}-2
as discussed above. However, for magnetoelectric media, the
spin-resolved emission is symmetric with respect to top ($\theta \in
[0,\pi/2)$) and bottom hemispheres ($\theta \in [\pi/2,\pi]$) whereas
  it is asymmetric for gyrotropic media. The asymmetry can be
  predicted based on the microscopic picture of underlying cyclotron
  motion. If the thermally or statistically averaged cyclotron motion
  of electrons is anti-clockwise in the $\bm{xy}$-plane (perpendicular
  gyrotropy axis) as viewed from $+\bm{z}$-direction, the resulting
  emission should be RCP along $+\bm{z}$-direction and LCP along
  $-\bm{z}$-direction.

We also point out that the material classes lead to unique dependence
of $S_3$ on the emission directions which is evident from the contour
plots. This can be informative from the perspective of classifying or
identifying materials using infrared imaging polarimetry in an actual
experiment.

{\bf Nonequilibrium torque}: We find that all gyrotropic planar slabs
experience a net nonzero torque along the direction of the gyrotropy
axis (examples 3 to 8) because of the overall loss of angular momentum
via emission of spin-polarized photons. For isotropic magnetoelectric
(Tellegen and Pasteur) media (examples 9,13), the total spectral
torque per unit area is zero which follows from the symmetry in top
and bottom hemispheres discussed above. For other materials, the net
torque is zero because of the cancellation upon integration over the
angle $\phi$. Note that we have focused on the planar slabs of
individual material classes for illustration.

\begin{figure*}[t!]
  \centering\includegraphics[width=\linewidth]{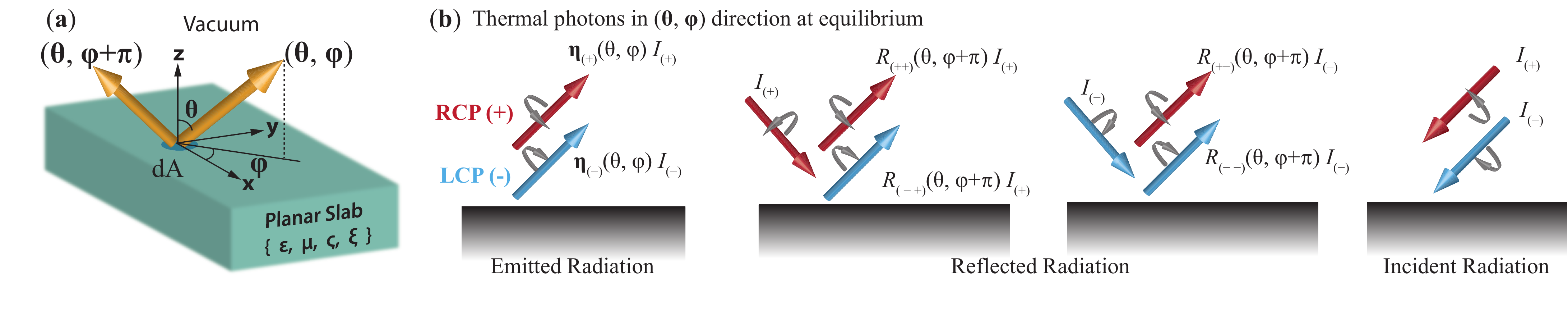}
  \caption{(a) The figure demonstrates the conventions used to
    describe the thermal radiation from the planar surface in a
    specific direction. The angles ($\theta,\phi$) characterize the
    emission direction (not the direction of incidence). Figure (b)
    illustrates the spectral radiance of thermal photons in
    ($\theta,\phi$) direction when the planar slab is at thermal
    equilibrium with the surrounding vacuum. These energy flux rates
    are described using the spin-resolved blackbody radiance
    ($I_{\pm}$), emissivities ($\eta_{\pm}$) and interconversion
    reflectances ($R$) for radiation incident along the conjugate
    ($\theta,\phi+\pi$) direction.  }
  \label{scm1}
\end{figure*}

For multilayered configurations combining different material classes,
a net nonzero torque can be obtained along a specific direction even
though it is absent for a single-layered slab. For instance, if the
isotropic magnetoelectric medium (examples 9, 13) is deposited on top
of another thin film of common dielectric or metallic material, a net
nonzero torque along $\bm{z}$-axis can be obtained because of the
asymmetric overall emission of spin-polarized photons in top and
bottom hemispheres. We provide these results for multilayered slabs in
the supplement. Based on a universal perspective, we identify material
classes required for engineering perpendicular and parallel components
of the nonequilibrium torque. We find that only materials satisfying
{\bf SKL-2} (examples 3,4,7,13) and isotropic reciprocal
magnetoelectric (Pasteur) media (example 9) can experience
\emph{perpendicular} nonequilibrium torque in a multilayered
configuration. Only materials satisfying {\bf SKL-3} (examples
5,6,8,15,16) and anisotropic reciprocal magnetoelectric media
(examples 11,12) can experience \emph{parallel} nonequilibrium torque
in a multilayered configuration.

{\bf Nonequilibrium force}: Figure~\ref{fig3} reveals that for a
single-layered geometry, only a reciprocal magnetoelectric planar slab
that causes cross coupling between $E_x-H_y$ and $E_y-H_x$ fields
(example 10), can lead to a nonzero nonequilibrium force along
$\bm{z}$-axis. For other materials, the symmetry of overall thermal
emission (summed over both spin states) in top and bottom hemispheres
leads to cancellation of the resulting force along $\bm{z}$-axis. The
figure also reveals that only planar slabs of anisotropic
nonreciprocal magnetoelectric materials (examples 15,16) can
experience nonequilibrium force parallel to the surface whose
direction is determined by the nature of the magneto-electric
coupling. If the magnetoelectric tensors ($\xib,\zetab$) indicate
($E_x-H_z$, $E_z-H_x$)-type cross coupling, the net force is along
$\bm{y}$-axis. If they indicate ($E_y-H_z$, $E_z-H_y$)-type cross
coupling, this force is along $\bm{x}$-axis. The multilayered slabs of
other material classes can also lead to nonequilibrium force as
discussed in the supplement. Based on a universal perspective, we find
that all material classes can experience a \emph{perpendicular}
nonequilibrium force in a multilayered configuration and only
materials satisfying {\bf SKL-3} can experience a \emph{parallel}
nonequilibrium force in a multilayered configuration. These results
indicate the possibility of engineering thermal-nonequilibrium
optomechanical forces and torques in planar geometries and identifying
or classifying materials based on experimentally measurable
characteristics.

\section{Discussion}
\label{analytics}

\subsection{Spin-resolved emissivities and absorptivities}
\label{derivation}

First, we derive the spin-resolved emissivities and absorptivities for
a semi-infinite half-space because of its simplicity and connection
with the fluctuational electrodynamic theory as explained in the
supplement. We then extend the derivation for a finite-thickness
planar slab. Since we assume linear material response and regular
reflection from the surface, the frequency $\omega$ and the angle
$\theta$ are not changed upon reflection. Hence, we can apply the
principle of total energy conservation for energy-exchange channels
characterized by $\omega,\theta$ separately and focus on the
polarization or spin-dependent properties. The flux rates are given by
equations~\ref{emission} and~\ref{absorption}.

\begin{figure*}[t!]
  \centering\includegraphics[width=\linewidth]{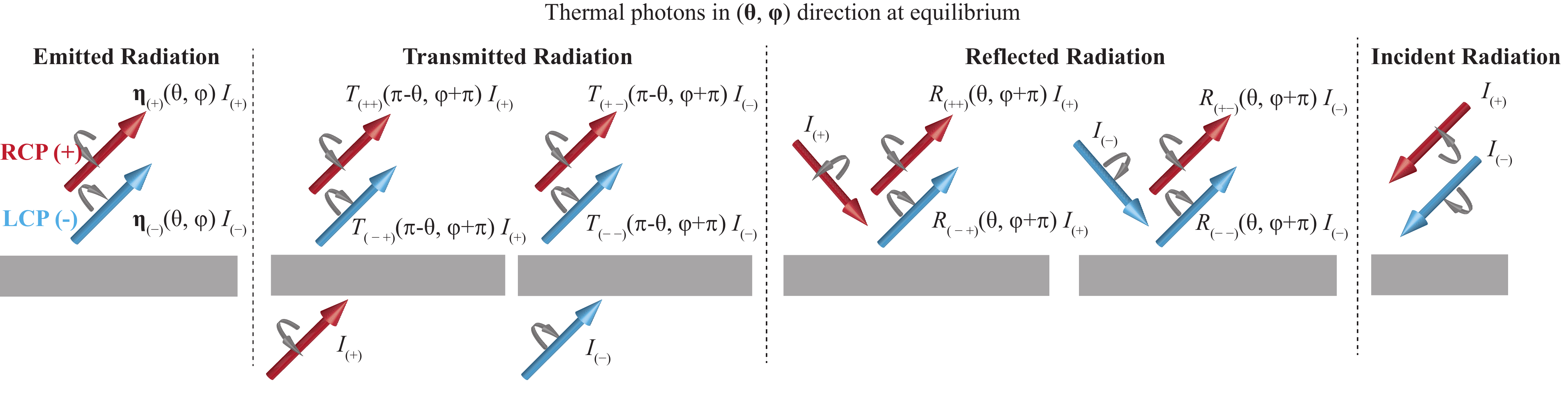}
  \caption{This figure illustrates the spectral radiance of thermal
    photons in ($\theta,\phi$) direction when the finite-thickness
    planar slab is at thermal equilibrium with the surrounding vacuum
    on both sides. These energy flux rates are described using the
    spin-resolved blackbody radiance ($I_{\pm}$), emissivities
    ($\eta_{\pm}$) and interconversion reflectances and
    transmittances. The emissivities are obtained by the equality of
    spin-resolved flux rates in opposite directions in the far-field
    under thermal equilibrium condition.}
  \label{scm2}
\end{figure*}

Figure \ref{scm1}(b) depicts the spin-resolved energy flux rates (RCP
and LCP radiation) at frequency $\omega$ in the direction
($\theta,\phi$) which contain emitted, reflected and incident
radiation. The fluxes are described at thermal equilibrium within the
radiometry paradigm~\cite{greffet1998field}. As shown in the rightmost
figure, the incident radiation contains both RCP ($I_{(+)}$) and LCP
($I_{(-)}$) radiation. Since the incident blackbody radiation is
unpolarized, it follows that $I_{(+)}=I_{(-)}=I_b/2$. The emitted
radiation is described using the spin-resolved emissivities as
$\eta_{(+)}I_{(+)}$ (RCP) and $\eta_{(-)}I_{(-)}$ (LCP). The reflected
radiation arises from the radiation incident along the conjugate
direction ($\theta,\phi+\pi$) and is described using the polarization
interconversion reflectances. For instance, as shown in the second
figure in \ref{scm1}(b), the incident RCP radiation ($I_{(+)}$) gets
reflected as $R_{(++)}(\theta,\phi+\pi)I_{(+)}$ (RCP) and
$R_{(-+)}(\theta,\phi+\pi)I_{(+)}$ (LCP) along ($\theta,\phi$)
direction. Note that we have characterized directions based on the
emitted radiation (indicated by arrows in (a)) and not the incident
radiation. This avoids the ambiguity in characterizing directions for
a thin film (finite-thickness slab described below) where transmitted
radiation is also taken into account. The interconversion reflectances
can be measured separately in an experiment or can be calculated for
the planar slab using the well-known Fresnel reflection coefficients
($r_{ss}$, $r_{sp}$, $r_{ps}$, $r_{pp}$) using the following
expressions (omitting the dependence on $\theta,\phi$ for brevity):
\begin{align}
  R_{(++/--)} = |(r_{ss}+r_{pp})\pm i(r_{sp}-r_{ps})|^2 / 4 \nonumber \\
  R_{(-+/+-)} = |(r_{ss}-r_{pp})\pm i(r_{sp}+r_{ps})|^2 / 4
\label{reflect}
\end{align}

The Fresnel reflection coefficient $r_{jk}$ for $j,k=[s,p]$ denotes
the amplitude of j-polarized reflected wave due to unit amplitude
k-polarized incident wave. Since the overall radiation in the
far-field is isotropic and unpolarized at thermal equilibrium, the
spin-resolved energy flux rates in opposite directions are equal. This
argument is justified by our previous work~\cite{khandekar2019spin}
which shows that the equilibrium spin angular momentum density and the
Poynting flux of thermal radiation are always zero in the far-field
although they can be nonzero in the near-field of certain
nonreciprocal media. By equating the incoming and outgoing
spin-resolved radiation flux rates in the far-field, we obtain the
following spin-resolved emissivities in terms of reflectances:
\begin{align}
\eta_{(+)}(\theta,\phi)=1-R_{(++)}(\theta,\phi+\pi)-
R_{(+-)}(\theta,\phi+\pi) \nonumber
\\ \eta_{(-)}(\theta,\phi)=1-R_{(--)}(\theta,\phi+\pi)-
R_{(-+)}(\theta,\phi+\pi)
\label{eta}
\end{align}

Similarly, it is straightforward to obtain the spin-resolved
absorptivities by considering the reflection of incident polarized
radiation $I_{(\pm)}$ separately. By subtracting the reflected flux
from the incident radiation flux, we obtain the following
spin-resolved absorptivities:
\begin{align}
\alpha_{(+)}(\theta,\phi)=1-R_{(++)}(\theta,\phi)-
R_{(-+)}(\theta,\phi) \nonumber
\\ \alpha_{(-)}(\theta,\phi)=1-R_{(--)}(\theta,\phi)-
R_{(+-)}(\theta,\phi)
\label{alpha}
\end{align}
We show in the supplement that the emissivities (Eq.\ref{eta}) can be
obtained within the scattering matrix formulation of fluctuational
electrodynamics theory. We further validate this derivation by proving
the thermodynamic consistency condition of net zero exchange of linear
and angular momentum of the planar slab with the environment at
thermal equilibrium. But before that, we extend this analysis to
obtain the spin-resolved emissivities and absorptivities for a planar
slab of finite thickness.

For a finite-thickness slab surrounded by vacuum on both sides, the
incident radiation, in addition to getting reflected, also gets
transmitted to the other side of the slab. It is straightforward to
obtain the emissivities in terms of reflectances and transmittances
using similar energy-balance considerations. In particular, as shown
in Fig.\ref{scm2}, the radiation in the direction ($\theta,\phi$)
contains emitted, transmitted, reflected and incident photons. The
transmitted radiation in ($\theta,\phi$) direction arises from the
radiation incident in the direction ($\pi-\theta,\phi+\pi$) on the
other side as depicted in Fig.\ref{scm2}(b). The directions are
consistently characterized based on the emission directions and not
the directions of the incidence or the associated wavevectors. The
transmittances are calculated from the associated Fresnel transmission
coefficients ($t_{ss}$, $t_{sp}$, $t_{ps}$, $t_{pp}$) using the
expressions (omitting the dependence on $\theta,\phi$ for brevity):
\begin{align}
  T_{(++/--)} = \frac{1}{4}|(t_{ss}+t_{pp})\pm i(t_{sp}-t_{ps})|^2 \nonumber \\
  T_{(-+/+-)} = \frac{1}{4}|(t_{ss}-t_{pp})\pm i(t_{sp}+t_{ps})|^2
\label{transmit}
\end{align}
The transmission coefficient $t_{jk}$ for $j,k=[s,p]$ denotes the
amplitude of j-polarized transmitted wave due to unit amplitude
k-polarized incident wave. Equating outgoing and incoming
spin-resolved flux rates in the opposite directions in the far-field,
we obtain the following expressions for the spin-resolved
emissivities:
\begin{align}
  \eta_{(+)}(\theta,\phi) &= 1 - R_{(+-)}(\theta,\phi+\pi) -
  R_{(++)}(\theta,\phi+\pi) \nonumber \\ &-
  T_{(+-)}(\pi-\theta,\phi+\pi) - T_{(++)}(\pi-\theta,\phi+\pi)
  \nonumber \\ \eta_{(-)}(\theta,\phi) &= 1 -
  R_{(-+)}(\theta,\phi+\pi) - R_{(--)}(\theta,\phi+\pi) \nonumber
  \\ &- T_{(-+)}(\pi-\theta,\phi+\pi) - T_{(--)}(\pi-\theta,\phi+\pi)
\label{etaT}
\end{align}
Similarly, the spin-resolved absorptivities are obtained by
subtracting the reflected and transmitted flux rates from the incident
polarized flux rates.
\begin{align}
  \alpha_{(+)}(\theta,\phi) &= 1 - R_{(-+)}(\theta,\phi) -
  R_{(++)}(\theta,\phi) \nonumber \\ &- T_{(-+)}(\theta,\phi) -
  T_{(++)}(\theta,\phi) \nonumber \\ \alpha_{(-)}(\theta,\phi) &= 1 -
  R_{(+-)}(\theta,\phi) - R_{(--)}(\theta,\phi) \nonumber \\ &-
  T_{(+-)}(\theta,\phi) - T_{(--)}(\theta,\phi)
\label{alphaT}
\end{align}

{\bf Detailed balance of linear and angular momentum exchange}: We now
demonstrate the theoretical consistency of the above derivation by
showing detailed balance of linear and angular momentum exchange
between the slab and its surroundings which must be maintained at
thermal equilibrium. Since angular momentum remains a largely
unexplored topic in the field of thermal
radiation~\cite{khandekar2019circularly,ott2018circular}, we explain
it in detail. A circularly polarized plane wave in vacuum normalized
to describe a single photon carries linear momentum of $\hbar\omega/c$
and angular momentum of $\pm \hbar$ along its direction of
propagation. It carries only spin angular momentum since the orbital
contributions are zero. Intrinsic orbital angular momentum is zero
because of the plane wavefront and extrinsic orbital angular momentum
is zero because of the cancellation over its infinite transverse
extent~\cite{barnett2016natures,barnett2017optical}. It also follows
from the spin angular momentum density expression [$\sim
  \Im\{\Ev^*(\omega) \times \Ev(\omega)\}$] that there are no
cross-polarization interaction terms in the net angular momentum when
the radiation consists of both RCP ($\ev_{(+)}$) and LCP ($\ev_{(-)}$)
photons. Therefore, similar to the energy flux, the angular momentum
flux of thermal photons can be considered separately for both
polarization states. Calculating the photon number flux rates using
Eqs.~\ref{emission} and~\ref{absorption}, we obtain both linear and
angular momentum flux rates for a finite-thickness planar slab. The
rates for a semi-infinite half-space are obtained by choosing zero
transmittances.

The rate of change of linear momentum $d\mathbf{p}/dt$ along
$+\mathbf{z}$-direction of the planar slab due to absorption,
reflection, and transmission of photons incident along ($\theta,\phi$)
direction and emitted by the slab along the same direction is given
below (abbreviating $dM=|cos\theta|d\omega d\Omega dA$ in
Eq.\ref{emission} for simplicity):
\begin{align}
\frac{dp_z}{dt}(\theta,\phi) &=
\bigg[\overbrace{-(\alpha_{(+)}+\alpha_{(-)})
    I_{b}(T_{\text{env}})}^{\text{absorption}} -
  \overbrace{(\eta_{(+)}+\eta_{(-)})I_b(T_0)}^{\text{emission}}
  \nonumber \\ &+ \sum_{j,k=(+,-)}
  \underbrace{-2R_{(jk)}I_b(T_\text{env})}_{\text{reflection}} \bigg]
\frac{1}{2c}\cos\theta dM
\label{dpzdt}
\end{align}
where $T_0$ is the temperature of the slab and $T_{\text{env}}$ is the
temperature of the environment.  The dependence of various quantities
on ($\omega,\theta,\phi$) is assumed and not mentioned above for
brevity. As an explanation of various terms, we consider incident RCP
(+) radiation in the top hemisphere and find the linear momentum
change of the slab along $+\mathbf{z}$-axis. It follows that
$\alpha_{(+)}$ portion gets absorbed imparting momentum $\propto
-\cos\theta\alpha_+ I_{b,T_{\text{env}}} \hbar\omega/c$,
$R_{(++)}+R_{(-+)}$ portion gets reflected in the positive
$\mathbf{z}$-direction imparting momentum $\propto
-2\cos\theta(R_{(++)}+R_{(-+)})I_{b,T_{\text{env}}} \hbar\omega/c$,
and $T_{(++)}+T_{(-+)}$ portion gets transmitted to the bottom
hemisphere without imparting any momentum to the slab. Because of the
emission of RCP radiation in the same direction, momentum $\propto
-\cos\theta\eta_{(+)}I_{b,T_0}$ is imparted to the slab. Negative sign
corresponds to the loss of the linear momentum by the slab.

Similarly, we obtain the rate of change of angular momentum
($d\mathbf{J}/dt$) of the planar slab whose $\mathbf{z}$-component is:
\begin{align}
\frac{dJ_z}{dt}(\theta,\phi) =
\bigg[\overbrace{-(\alpha_{(+)}-\alpha_{(-)})
    I_{b,T_{\text{env}}}}^{\text{absorption}} -
  \overbrace{(\eta_{(+)}-\eta_{(-)})I_{b,T_0}}^{\text{emission}}
  \nonumber \\ - 2(\underbrace{R_{(++)}-R_{(--)}}_{\text{reflection}}
  +
  \underbrace{T_{(-+)}-T_{(+-)}}_{\text{transmission}})I_{b,T_\text{env}}
  \bigg] \frac{1}{2\omega}\cos\theta dM
\label{dJzdt}
\end{align}
The $\mathbf{x}$-components of both linear and angular momentum
transfer rates are:
\begin{align}
  \label{dpxdt}
\frac{dp_x}{dt}(\theta,\phi) &=
\bigg[-(\alpha_{(+)}+\alpha_{(-)})I_{b,T_{\text{env}}} \nonumber
  \\ &-(\eta_{(+)}+\eta_{(-)})I_{b,T_0}\bigg]\frac{1}{2c}
\sin\theta\cos\phi dM \\ \frac{dJ_x}{dt}(\theta,\phi) &=
\bigg[-(\alpha_{(+)}-\alpha_{(-)})I_{b,T_{\text{env}}}-(\eta_{(+)}
  -\eta_{(-)})I_{b,T_0} \nonumber
  \\ &\hspace{-45pt}-2(R_{(-+)}-R_{(+-)}+T_{(-+)}-T_{(+-)})I_{b,T_{\text{env}}}\bigg]\frac{\sin\theta\cos\phi dM}{2\omega}
\label{dJxdt}
\end{align}
The $\mathbf{y}$-components are obtained by replacing $\cos\phi$ with
$\sin\phi$ in above two expressions. It is then straightforward to
calculate the spectral force per unit area ($\mathbf{F}_s$) and the
spectral torque per unit area ($\bm{\tau}_s$) experienced by the
planar slab due to the thermal emission of photons of frequency
$\omega$:
\begin{align}
\label{Fsa}
  \mathbf{F}_s=\frac{d\mathbf{F}}{d\omega dA}(\omega) = \int
\frac{d\mathbf{p}}{dt d\omega dA}(\omega,\theta,\phi)
\\ \bm{\tau}_s=\frac{d\bm{\tau}_s}{d\omega dA}(\omega) = \int
\frac{d\bm{J}}{dt d\omega dA}(\omega,\theta,\phi)
\label{Tsa}
\end{align}
Here the integration is over the solid angle $d\Omega$ and the terms
in the denominator $d\omega dA$ come from $dM$ introduced in
Eqs~\ref{dpzdt},\ref{dJzdt},\ref{dpxdt},\ref{dJxdt}. The final
expressions are integrated over the frequency $\omega$ for calculating
total force and total torque per unit area.

We now substitute the spin-resolved emissivities (Eq.\ref{etaT}) and
absorptivities (Eq.\ref{alphaT}) in the above expressions and compute
the rates of linear and angular momentum transfer under thermal
equilibrium condition ($T_0=T_{\text{env}}$). We find that the
following vectorial equalities always hold true:
\begin{align}
\frac{d\mathbf{\bm{p}}}{dt}(\theta,\phi) &+
\frac{d\mathbf{\bm{p}}}{dt}(\theta,\phi+\pi) \nonumber \\ &+
\frac{d\mathbf{\bm{p}}}{dt}(\pi-\theta,\phi)+
\frac{d\mathbf{\bm{p}}}{dt}(\pi-\theta,\phi+\pi)=0 \nonumber
\\ \frac{d\bm{J}}{dt}(\theta,\phi) &+
\frac{d\bm{J}}{dt}(\theta,\phi+\pi) \nonumber \\ &+
\frac{d\bm{J}}{dt}(\pi-\theta,\phi)+
\frac{d\bm{J}}{dt}(\pi-\theta,\phi+\pi)=0
\label{db}
\end{align}
It follows upon integration over all emission directions that there is
no net force or torque on the planar slab which is at thermal
equilibrium with the surrounding environment. Since the spin-resolved
emissivities (Eq.\ref{etaT}) are derived without invoking the concept
of electromagnetic reciprocity, and are new to the best of our
knowledge, we note that the above analysis of detailed balance of
linear and angular momentum transfer provides a thermodynamic
consistency of the derivation. The above derivation is also crucial
for computing the fluctuations-induced force and torque on the planar
slabs under the nonequilibrium condition when $T_{\text{env}} \neq
T_{0}$ as discussed above.


\subsection{Derivation of Spin-resolved Kirchhoff's laws}

The calculation of spin-resolved emissivities and absorptivities
requires the calculation of Fresnel reflection and transmission
coefficients. While this is a very well-known topic in the literature,
we note that, apart from the trivial isotropic media, the closed-form
semi-analytic expressions of reflection coefficients are possible only
for semi-infinite half-spaces of very few bianisotropic material
classes~\cite{graglia1991reflection,tretyakov1994reflection,
  mueller1971reflection}. We separately show this calculation in the
supplement for a gyromagnetic medium. However, the semi-analytic
approach is inadequate because we are interested in finding the laws
which are applicable for a generic case of a finite-thickness slab
that can possibly contain multiple layers of various materials. It can
also be a composite medium where a single layer can simultaneously
exhibit response of multiple material classes. Since meaningful
closed-form expressions are not possible for these complicated cases,
we have developed numerical tools to calculate the reflection and
transmission coefficients in an exact manner as discussed in the
methods section. We have made them open-source\footnote{This software
  is available on GitHub \url{https://github.com/chinmayCK/Fresnel}
  under MIT license} so that the reader can reproduce and verify the
results in this work and also use these versatile tools for any other
research activities. Our computer-assisted but exact approach for the
discovery of Kirchhoff's laws is similar in spirit to that of an
increasing use of machine learning, optimization, and inverse design
computational techniques in many scientific disciplines for making
fundamental discoveries.

{\bf Reciprocal materials}: For a planar slab of reciprocal media, we
find that the reflection coefficients satisfy the relations
$r_{ss,pp}(\theta,\phi)=r_{ss,pp}(\theta,\phi+\pi)$ and
$r_{sp}(\theta,\phi)=-r_{ps}(\theta,\phi+\pi)$. The transmission
coefficients satisfy
$t_{ss,pp}(\theta,\phi)=t_{ss,pp}(\pi-\theta,\phi+\pi)$ and
$t_{sp}(\theta,\phi)=-t_{ps}(\pi-\theta,\phi+\pi)$. These relations
can be inferred from the Green's function on the vacuum side of the
geometry~\cite{bimonte2007general}. Substituting these relations in
Eqs.~\ref{etaT} and \ref{alphaT}, we find that the spin-resolved
emissivity is equal to the spin-resolved absorptivity for reciprocal
media. $\eta_{\pm}(\omega,\theta,\phi)=\alpha_{\pm}(\omega,\theta,\phi)$.
This is {\bf SKL-1} given in Eq.~\ref{KL1}. For a chiral absorber
$\alpha_{(+)} \neq \alpha_{(-)}$ from which it follows that
$\eta_{(+)} \neq \eta_{(-)}$. Therefore, a chiral absorber can emit
circularly polarized thermal radiation. But this well-known law is
valid only for reciprocal media. Interestingly, for nonreciprocal
media, we find other formulations of this spin-resolved Kirchhoff's
law.

{\bf Nonreciprocal materials}: For gyrotropic materials with the
gyrotropy axis perpendicular to the planar surface and for
nonreciprocal isotropic magneto-electric materials (Tellegen media)
having diagonal $\xib,\zetab$ tensors, we find that the reflection
coefficients satisfy the relations
$r_{ss,pp}(\theta,\phi)=r_{ss,pp}(\theta,\phi+\pi)$ and
$r_{sp}(\theta,\phi)=r_{ps}(\theta,\phi+\pi)$. The transmission
coefficients satisfy the relations
$t_{ss,pp}(\theta,\phi)=t_{ss,pp}(\pi-\theta,\phi+\pi)$ and
$t_{sp}(\theta,\phi)=t_{ps}(\pi-\theta,\phi+\pi)$. Substituting these
relations in Eqs.~\ref{etaT} and \ref{alphaT}, we obtain
$\eta_{\pm}(\omega,\theta,\phi)=\alpha_{\mp}(\omega,\theta,\phi)$.
This is {\bf SKL-2} given in Eq.~\ref{KL2}. 

Interestingly, multilayered planar slabs of other nonreciprocal
gyrotopic and anisotropic magneto-electric materials also exhibit
simplifying relations between the Fresnel coefficients. In particular,
if the magnetoelectric coupling tensors ($\xib,\zetab$) indicate cross
coupling between electric and magnetic fields lying perpendicular and
parallel to the slab ($E_x-H_z$, $E_z-H_x$, $E_y-H_z$ or $E_z-H_y$) or
if the medium is gyrotropic with the gyrotropy axis parallel to the
surface, the reflection coefficients satisfy the condition
$r_{sp}(\theta,\phi)=-r_{ps}(\theta,\phi)$ and the transmission
coefficients satisfy $t_{ss,pp}(\pi-\theta,\phi+\pi)=
t_{ss,pp}(\theta,\phi+\pi)$ and
$t_{sp}(\pi-\theta,\phi+\pi)=-t_{ps}(\theta,\phi+\pi)$. Substituting
these relations in Eqs.~\ref{etaT} and \ref{alphaT}, we obtain
$\eta_{\pm}(\omega,\theta,\phi)=\alpha_{\pm}(\omega,\theta,\phi+\pi)$.
This is {\bf SKL-3} given in Eq.~\ref{KL3}. We note that the above
underlying relations are not obvious and their universal applicability
for multilayered or composite configurations of many bianisotropic
material classes has not been reported before. The explanation of
their specific form based on the underlying structural characteristics
or symmetries of material classes is a nontrivial problem that
motivates new analogies or methods~\cite{padilla2007group} across a
broad set of materials.


\section{Concluding Remarks}

We consider a different perspective of spin-resolved or spin-polarized
thermal radiation and analyze it comprehensively for reciprocal and
nonreciprocal media. We use radiometry paradigm to derive the
spin-resolved emissivity and absorptivity without invoking the concept
of electromagnetic reciprocity. We further provide a validation of
this derivation by showing the detailed balance of both linear and
angular momentum transfer rates when the planar slab is at thermal
equilibrium with its environment. The spin-resolved expressions ensure
that there is no net nonzero force or torque on the planar slab which
is a necessary condition to maintain thermal equilibrium with the
environment. Our emphasis on these theory details and derivation is
important because, although thermal radiation is a very old topic, its
extension to nonreciprocal media is a recent development and reviving
the old radiometry paradigm in that context (uncommon in recent works)
requires careful analysis. Besides, the concept of detailed balance of
angular momentum (in addition to energy) provides a fundamentally new
insight related to the interplay of photon spin and thermal
radiation. This concept can be extended to analyze heat transfer at
thermal equilibrium in many other systems for validating underlying
physical models or deriving similar heat-transfer laws.

Through a universal perspective of spin-resolved thermal emission for
generic bianisotropic material classes, we discover three useful
spin-resolved Kirchhoff's laws of thermal radiation applicable for
reciprocal and nonreciprocal planar media. While the first law
applicable for the reciprocal media ({\bf SKL}-1) is known, the
remaining two laws for the nonreciprocal planar media are new. We
emphasize that these laws are applicable for composite media or
multilayered configurations of many material classes, for all
frequencies, emission directions, geometry and material
parameters. Since Kirchhoff's laws are at the foundation of the field
of thermal radiation and are applicable only for reciprocal media,
their generalization to nonreciprocal media is a very important
fundamental result.
  
We also propose an experiment to verify the spin-resolved Kirchhoff's
laws conveniently by using a single material system. We note that we
can obtain another formulation of directional Kirchhoff's laws
relating total emissivity ($\eta=\eta_{(+)}+\eta_{(-)}$) with total
absorptivity ($\alpha=\alpha_{(+)}+\alpha_{(-)}$) by summing over both
spin polarization states. The spin-resolved Kirchhoff's laws and their
derivatives will be useful for optimizing directional radiative heat
transfer~\cite{green2012time,zhu2014near} and engineering broadband
circularly polarized thermal emission~\cite{wadsworth2011broadband}.

While the calculation of Fresnel coefficients for specific cases of
bianisotropic media is a well-known topic in the literature, a
universal perspective revealing the interesting underlying relations
in this work is new to the best of our knowledge in that research
area. They can be potentially used for discovering similar laws
concerning different (not necessarily thermal) radiation
properties. Moreover, we emphasize that this work is not merely the
derivation of Fresnel coefficients. The central results are not
obvious from the existing literature on Fresnel coefficients for
bianisotropic media, since they primarily depend on the important
theory details and derivation provided in this work. We also note that
the concepts like spin-resolved radiative heat transfer, bianisotropic
media, photon spin and angular momentum remain largely unexplored in
the field of thermal radiation (see recent
reviews~\cite{li2018nanophotonic}). Our work simultaneously advances
these new frontiers. We believe that these inquiries will be
insightful for other research areas. For example, angular momentum
transfer via spin-polarized thermal radiation (a different perspective
introduced here and in our previous work~\cite{khandekar2019spin}) can
be relevant in the separate field of spintronics.

In addition to the spin-resolved Kirchhoff's laws, we also show other
thermal spin photonic features for several material classes. First, we
show a striking result that the planar media of many materials can
emit spin-polarized thermal light. This includes all nonreciprocal
media as well as reciprocal media such as uni/bi-axial anisotropic
crystals and reciprocal magnetoelectric (chiral) media. Second, we
demonstrate nonequilibrium force and torque experienced by planar
slabs due to the loss of linear and angular momentum respectively, via
thermal radiation into the environment at lower temperature. We also
identify the material classes that can be used for engineering
parallel and perpendicular components of nonequilibrium force and
torque in multilayered planar slabs. These findings will be useful
from the perspective of practical applications such as designing
directional and spin-polarized light sources, engineering thermal
optomechanical forces and torques, and classifying or identifying
materials using infrared imaging polarimetry.

\section{Methods}

The calculation of Fresnel reflection and transmission coefficients
for the planar media involves two steps. The first step requires
calculation of the wavevectors and the plane-wave solutions of the
electromagnetic fields inside each layer of the planar slab. We
consider most general multi-layered geometry. The second step
involves the calculation of these coefficients by enforcing the
boundary conditions at each interface for incidence of either
$\ev_{s}$ or $\ev_{p}$ polarized light.

A generic, homogeneous bianisotropic medium is described using the
constitutive relations given in Eq.\ref{consti}. By writing
electromagnetic fields inside the material as $[\Ev,
  \sqrt{\frac{\mu_0}{\varepsilon_0}}\Hv]^T
e^{i(\kv_{\parallel}\cdot\Rv+ k_z z -i\omega t)}$ and using these
constitutive relations in Maxwell's equations, we obtain the following
dimensionless dispersion equation for waves inside the material:
\begin{align}
  &\text{det}(M+M_k) = 0, \hspace{5pt} \text{for} \hspace{5pt}
  M=\begin{bmatrix} \epsb & \xib \\ \zetab & \mub \end{bmatrix},
  M_k=\begin{bmatrix} 0 & \kb/k_0 \\ -\kb/k_0 & 0 \end{bmatrix}
  \nonumber \\ &\kb=\begin{bmatrix} 0 & -k_z & k_\parallel\sin\phi
  \\ k_z & 0 & -k_\parallel\cos\phi \\ -k_\parallel\sin\phi &
  k_\parallel\cos\phi & 0 \end{bmatrix}
\end{align}
Here, $6\times 6$ material tensor $M$ describes the constitutive
relations and $M_k$ corresponds to the curl operator acting on plane
waves. $\kv_\parallel$ is the parallel component of the incident wave
which is conserved at each interface and hence is the same for all
layers. $k_z$ is the $\bm{z}$-component of the wavevector which is
obtained by solving the above equation for given
$k_\parallel=|\kv_\parallel|$, the angle $\phi$ made by
$\kv_\parallel$ with the $\bm{x}$-axis, and the material parameters
$\epsb,\mub,\xib,\zetab$. The associated fields are the null-space
eigenstates of the above dispersion equation. For all media, there are
four solutions inside each layer of the planar slab where two
solutions correspond to waves going in $+\bm{z}$ direction and the
remaining two correspond to waves going in $-\bm{z}$ direction.

We now consider the incidence of $\ev_s$ or $\ev_p$ polarized light.
In vacuum, the polarization vectors $\ev_{j\pm}$ for $j={s,p}$ with
$\pm$ denoting waves going along $\pm\bm{z}$ directions are:
\begin{align}
\ev_{s\pm}=\begin{bmatrix}\sin\phi \\ -\cos\phi \\ 0 \end{bmatrix},
\ev_{p\pm}=\frac{-1}{k_0} \begin{bmatrix}\pm k_z\cos\phi \\ \pm
  k_z\sin\phi \\ -k_\parallel \end{bmatrix}
\label{spvectors}
\end{align}
Thus, for an incident $\ev_{s-}$ polarized light (in the top
hemisphere for a planar slab) of unit amplitude, the reflected fields
in the same layer are $r_{ss}\ev_{s+} + r_{ps}\ev_{p+}$ and the
transmitted fields on the other side of the slab are
$t_{ss}\ev_{s-}+t_{ps}\ev_{p-}$. Inside each layer, the fields are
described in the basis of four (null space) eigenstates using four
unknowns. Thus, for a planar slab of $N$ layers, there are $4N+4$
unknowns including the four reflection and transmission
coefficients. These can be readily calculated by using $4(N+1)$
boundary conditions (continuity of $E_x,E_y,H_x,H_y$ fields) for $N+1$
interfaces of the planar geometry. The computational tools to calculate
these coefficients are made available on GitHub
\url{https://github.com/chinmayCK/Fresnel} under MIT license. 

\section{Acknowledgments}
This work was supported by the US Department of Energy, Office of
Basic Energy Science under award number DE-SC0017717, DARPA Nascent
Light-Matter Interaction program, and the Lillian Gilbreth
Postdoctoral Fellowship program at Purdue University (CK).

\bibliography{photon}

\begin{thebibliography}{49}%
\makeatletter
\providecommand \@ifxundefined [1]{%
 \@ifx{#1\undefined}
}%
\providecommand \@ifnum [1]{%
 \ifnum #1\expandafter \@firstoftwo
 \else \expandafter \@secondoftwo
 \fi
}%
\providecommand \@ifx [1]{%
 \ifx #1\expandafter \@firstoftwo
 \else \expandafter \@secondoftwo
 \fi
}%
\providecommand \natexlab [1]{#1}%
\providecommand \enquote  [1]{``#1''}%
\providecommand \bibnamefont  [1]{#1}%
\providecommand \bibfnamefont [1]{#1}%
\providecommand \citenamefont [1]{#1}%
\providecommand \href@noop [0]{\@secondoftwo}%
\providecommand \href [0]{\begingroup \@sanitize@url \@href}%
\providecommand \@href[1]{\@@startlink{#1}\@@href}%
\providecommand \@@href[1]{\endgroup#1\@@endlink}%
\providecommand \@sanitize@url [0]{\catcode `\\12\catcode `\$12\catcode
  `\&12\catcode `\#12\catcode `\^12\catcode `\_12\catcode `\%12\relax}%
\providecommand \@@startlink[1]{}%
\providecommand \@@endlink[0]{}%
\providecommand \url  [0]{\begingroup\@sanitize@url \@url }%
\providecommand \@url [1]{\endgroup\@href {#1}{\urlprefix }}%
\providecommand \urlprefix  [0]{URL }%
\providecommand \Eprint [0]{\href }%
\providecommand \doibase [0]{http://dx.doi.org/}%
\providecommand \selectlanguage [0]{\@gobble}%
\providecommand \bibinfo  [0]{\@secondoftwo}%
\providecommand \bibfield  [0]{\@secondoftwo}%
\providecommand \translation [1]{[#1]}%
\providecommand \BibitemOpen [0]{}%
\providecommand \bibitemStop [0]{}%
\providecommand \bibitemNoStop [0]{.\EOS\space}%
\providecommand \EOS [0]{\spacefactor3000\relax}%
\providecommand \BibitemShut  [1]{\csname bibitem#1\endcsname}%
\let\auto@bib@innerbib\@empty
\bibitem [{\citenamefont {Wu}\ \emph {et~al.}(2014)\citenamefont {Wu},
  \citenamefont {Arju}, \citenamefont {Kelp}, \citenamefont {Fan},
  \citenamefont {Dominguez}, \citenamefont {Gonzales}, \citenamefont {Tutuc},
  \citenamefont {Brener},\ and\ \citenamefont {Shvets}}]{wu2014spectrally}%
  \BibitemOpen
  \bibfield  {author} {\bibinfo {author} {\bibfnamefont {C.}~\bibnamefont
  {Wu}}, \bibinfo {author} {\bibfnamefont {N.}~\bibnamefont {Arju}}, \bibinfo
  {author} {\bibfnamefont {G.}~\bibnamefont {Kelp}}, \bibinfo {author}
  {\bibfnamefont {J.~A.}\ \bibnamefont {Fan}}, \bibinfo {author} {\bibfnamefont
  {J.}~\bibnamefont {Dominguez}}, \bibinfo {author} {\bibfnamefont
  {E.}~\bibnamefont {Gonzales}}, \bibinfo {author} {\bibfnamefont
  {E.}~\bibnamefont {Tutuc}}, \bibinfo {author} {\bibfnamefont
  {I.}~\bibnamefont {Brener}}, \ and\ \bibinfo {author} {\bibfnamefont
  {G.}~\bibnamefont {Shvets}},\ }\href
  {https://www.nature.com/articles/ncomms4892} {\bibfield  {journal} {\bibinfo
  {journal} {Nat. Commun.}\ }\textbf {\bibinfo {volume} {5}},\ \bibinfo {pages}
  {3892} (\bibinfo {year} {2014})}\BibitemShut {NoStop}%
\bibitem [{\citenamefont {Yin}\ \emph {et~al.}(2013)\citenamefont {Yin},
  \citenamefont {Schaferling}, \citenamefont {Metzger},\ and\ \citenamefont
  {Giessen}}]{yin2013interpreting}%
  \BibitemOpen
  \bibfield  {author} {\bibinfo {author} {\bibfnamefont {X.}~\bibnamefont
  {Yin}}, \bibinfo {author} {\bibfnamefont {M.}~\bibnamefont {Schaferling}},
  \bibinfo {author} {\bibfnamefont {B.}~\bibnamefont {Metzger}}, \ and\
  \bibinfo {author} {\bibfnamefont {H.}~\bibnamefont {Giessen}},\ }\href
  {https://pubs.acs.org/doi/10.1021/nl403705k} {\bibfield  {journal} {\bibinfo
  {journal} {Nano Lett.}\ }\textbf {\bibinfo {volume} {13}},\ \bibinfo {pages}
  {6238} (\bibinfo {year} {2013})}\BibitemShut {NoStop}%
\bibitem [{\citenamefont {Dyakov}\ \emph {et~al.}(2018)\citenamefont {Dyakov},
  \citenamefont {Semenenko}, \citenamefont {Gippius},\ and\ \citenamefont
  {Tikhodeev}}]{dyakov2018magnetic}%
  \BibitemOpen
  \bibfield  {author} {\bibinfo {author} {\bibfnamefont {S.~A.}\ \bibnamefont
  {Dyakov}}, \bibinfo {author} {\bibfnamefont {V.~A.}\ \bibnamefont
  {Semenenko}}, \bibinfo {author} {\bibfnamefont {N.~A.}\ \bibnamefont
  {Gippius}}, \ and\ \bibinfo {author} {\bibfnamefont {S.~G.}\ \bibnamefont
  {Tikhodeev}},\ }\href {\doibase 10.1103/PhysRevB.98.235416} {\bibfield
  {journal} {\bibinfo  {journal} {Phys. Rev. B}\ }\textbf {\bibinfo {volume}
  {98}},\ \bibinfo {pages} {235416} (\bibinfo {year} {2018})}\BibitemShut
  {NoStop}%
\bibitem [{\citenamefont {Shitrit}\ \emph {et~al.}(2013)\citenamefont
  {Shitrit}, \citenamefont {Yulevich}, \citenamefont {Maguid}, \citenamefont
  {Ozeri}, \citenamefont {Veksler}, \citenamefont {Kleiner},\ and\
  \citenamefont {Hasman}}]{shitrit2013spin}%
  \BibitemOpen
  \bibfield  {author} {\bibinfo {author} {\bibfnamefont {N.}~\bibnamefont
  {Shitrit}}, \bibinfo {author} {\bibfnamefont {I.}~\bibnamefont {Yulevich}},
  \bibinfo {author} {\bibfnamefont {E.}~\bibnamefont {Maguid}}, \bibinfo
  {author} {\bibfnamefont {D.}~\bibnamefont {Ozeri}}, \bibinfo {author}
  {\bibfnamefont {D.}~\bibnamefont {Veksler}}, \bibinfo {author} {\bibfnamefont
  {V.}~\bibnamefont {Kleiner}}, \ and\ \bibinfo {author} {\bibfnamefont
  {E.}~\bibnamefont {Hasman}},\ }\href
  {https://science.sciencemag.org/content/340/6133/724} {\bibfield  {journal}
  {\bibinfo  {journal} {Science}\ }\textbf {\bibinfo {volume} {340}},\ \bibinfo
  {pages} {724} (\bibinfo {year} {2013})}\BibitemShut {NoStop}%
\bibitem [{\citenamefont {Wadsworth}\ \emph {et~al.}(2011)\citenamefont
  {Wadsworth}, \citenamefont {Clem}, \citenamefont {Branson},\ and\
  \citenamefont {Boreman}}]{wadsworth2011broadband}%
  \BibitemOpen
  \bibfield  {author} {\bibinfo {author} {\bibfnamefont {S.}~\bibnamefont
  {Wadsworth}}, \bibinfo {author} {\bibfnamefont {P.}~\bibnamefont {Clem}},
  \bibinfo {author} {\bibfnamefont {E.}~\bibnamefont {Branson}}, \ and\
  \bibinfo {author} {\bibfnamefont {G.}~\bibnamefont {Boreman}},\ }\href
  {https://www.osapublishing.org/ome/abstract.cfm?uri=ome-1-3-466} {\bibfield
  {journal} {\bibinfo  {journal} {Opt. Mater. Exp.}\ }\textbf {\bibinfo
  {volume} {1}},\ \bibinfo {pages} {466} (\bibinfo {year} {2011})}\BibitemShut
  {NoStop}%
\bibitem [{\citenamefont {Miller}\ \emph {et~al.}(2017)\citenamefont {Miller},
  \citenamefont {Zhu},\ and\ \citenamefont {Fan}}]{miller2017universal}%
  \BibitemOpen
  \bibfield  {author} {\bibinfo {author} {\bibfnamefont {D.}~\bibnamefont
  {Miller}}, \bibinfo {author} {\bibfnamefont {L.}~\bibnamefont {Zhu}}, \ and\
  \bibinfo {author} {\bibfnamefont {S.}~\bibnamefont {Fan}},\ }\href
  {https://www.pnas.org/content/114/17/4336} {\bibfield  {journal} {\bibinfo
  {journal} {Proc. Natl. Acad. Sci.}\ }\textbf {\bibinfo {volume} {114}},\
  \bibinfo {pages} {4336} (\bibinfo {year} {2017})}\BibitemShut {NoStop}%
\bibitem [{\citenamefont {Zhu}\ and\ \citenamefont {Fan}(2014)}]{zhu2014near}%
  \BibitemOpen
  \bibfield  {author} {\bibinfo {author} {\bibfnamefont {L.}~\bibnamefont
  {Zhu}}\ and\ \bibinfo {author} {\bibfnamefont {S.}~\bibnamefont {Fan}},\
  }\href {\doibase 10.1103/PhysRevB.90.220301} {\bibfield  {journal} {\bibinfo
  {journal} {Phys. Rev. B}\ }\textbf {\bibinfo {volume} {90}},\ \bibinfo
  {pages} {220301} (\bibinfo {year} {2014})}\BibitemShut {NoStop}%
\bibitem [{\citenamefont {Moncada-Villa}\ \emph {et~al.}(2015)\citenamefont
  {Moncada-Villa}, \citenamefont {Fern\'andez-Hurtado}, \citenamefont
  {Garc\'{\i}a-Vidal}, \citenamefont {Garc\'{\i}a-Mart\'{\i}n},\ and\
  \citenamefont {Cuevas}}]{moncada2015magnetic}%
  \BibitemOpen
  \bibfield  {author} {\bibinfo {author} {\bibfnamefont {E.}~\bibnamefont
  {Moncada-Villa}}, \bibinfo {author} {\bibfnamefont {V.}~\bibnamefont
  {Fern\'andez-Hurtado}}, \bibinfo {author} {\bibfnamefont {F.~J.}\
  \bibnamefont {Garc\'{\i}a-Vidal}}, \bibinfo {author} {\bibfnamefont
  {A.}~\bibnamefont {Garc\'{\i}a-Mart\'{\i}n}}, \ and\ \bibinfo {author}
  {\bibfnamefont {J.~C.}\ \bibnamefont {Cuevas}},\ }\href {\doibase
  10.1103/PhysRevB.92.125418} {\bibfield  {journal} {\bibinfo  {journal} {Phys.
  Rev. B}\ }\textbf {\bibinfo {volume} {92}},\ \bibinfo {pages} {125418}
  (\bibinfo {year} {2015})}\BibitemShut {NoStop}%
\bibitem [{\citenamefont {Latella}\ and\ \citenamefont
  {Ben-Abdallah}(2017)}]{latella2017giant}%
  \BibitemOpen
  \bibfield  {author} {\bibinfo {author} {\bibfnamefont {I.}~\bibnamefont
  {Latella}}\ and\ \bibinfo {author} {\bibfnamefont {P.}~\bibnamefont
  {Ben-Abdallah}},\ }\href {\doibase 10.1103/PhysRevLett.118.173902} {\bibfield
   {journal} {\bibinfo  {journal} {Phys. Rev. Lett.}\ }\textbf {\bibinfo
  {volume} {118}},\ \bibinfo {pages} {173902} (\bibinfo {year}
  {2017})}\BibitemShut {NoStop}%
\bibitem [{\citenamefont {Ben-Abdallah}(2016)}]{ben2016photon}%
  \BibitemOpen
  \bibfield  {author} {\bibinfo {author} {\bibfnamefont {P.}~\bibnamefont
  {Ben-Abdallah}},\ }\href {\doibase 10.1103/PhysRevLett.116.084301} {\bibfield
   {journal} {\bibinfo  {journal} {Phys. Rev. Lett.}\ }\textbf {\bibinfo
  {volume} {116}},\ \bibinfo {pages} {084301} (\bibinfo {year}
  {2016})}\BibitemShut {NoStop}%
\bibitem [{\citenamefont {Herz}\ and\ \citenamefont
  {Biehs}(2019)}]{herz2019green}%
  \BibitemOpen
  \bibfield  {author} {\bibinfo {author} {\bibfnamefont {F.}~\bibnamefont
  {Herz}}\ and\ \bibinfo {author} {\bibfnamefont {S.-A.}\ \bibnamefont
  {Biehs}},\ }\href
  {https://iopscience.iop.org/article/10.1209/0295-5075/127/44001/meta}
  {\bibfield  {journal} {\bibinfo  {journal} {EPL}\ }\textbf {\bibinfo {volume}
  {127}},\ \bibinfo {pages} {44001} (\bibinfo {year} {2019})}\BibitemShut
  {NoStop}%
\bibitem [{\citenamefont {Abraham~Ekeroth}\ \emph {et~al.}(2017)\citenamefont
  {Abraham~Ekeroth}, \citenamefont {Garc\'{\i}a-Mart\'{\i}n},\ and\
  \citenamefont {Cuevas}}]{ekeroth2017thermal}%
  \BibitemOpen
  \bibfield  {author} {\bibinfo {author} {\bibfnamefont {R.~M.}\ \bibnamefont
  {Abraham~Ekeroth}}, \bibinfo {author} {\bibfnamefont {A.}~\bibnamefont
  {Garc\'{\i}a-Mart\'{\i}n}}, \ and\ \bibinfo {author} {\bibfnamefont {J.~C.}\
  \bibnamefont {Cuevas}},\ }\href {\doibase 10.1103/PhysRevB.95.235428}
  {\bibfield  {journal} {\bibinfo  {journal} {Phys. Rev. B}\ }\textbf {\bibinfo
  {volume} {95}},\ \bibinfo {pages} {235428} (\bibinfo {year}
  {2017})}\BibitemShut {NoStop}%
\bibitem [{\citenamefont {Zhu}\ \emph {et~al.}(2018)\citenamefont {Zhu},
  \citenamefont {Guo},\ and\ \citenamefont {Fan}}]{zhu2018theory}%
  \BibitemOpen
  \bibfield  {author} {\bibinfo {author} {\bibfnamefont {L.}~\bibnamefont
  {Zhu}}, \bibinfo {author} {\bibfnamefont {Y.}~\bibnamefont {Guo}}, \ and\
  \bibinfo {author} {\bibfnamefont {S.}~\bibnamefont {Fan}},\ }\href {\doibase
  10.1103/PhysRevB.97.094302} {\bibfield  {journal} {\bibinfo  {journal} {Phys.
  Rev. B}\ }\textbf {\bibinfo {volume} {97}},\ \bibinfo {pages} {094302}
  (\bibinfo {year} {2018})}\BibitemShut {NoStop}%
\bibitem [{\citenamefont {\ifmmode \check{Z}\else
  \v{Z}\fi{}uti\ifmmode~\acute{c}\else \'{c}\fi{}}\ \emph
  {et~al.}(2004)\citenamefont {\ifmmode \check{Z}\else
  \v{Z}\fi{}uti\ifmmode~\acute{c}\else \'{c}\fi{}}, \citenamefont {Fabian},\
  and\ \citenamefont {Das~Sarma}}]{RevModPhys.76.323}%
  \BibitemOpen
  \bibfield  {author} {\bibinfo {author} {\bibfnamefont {I.}~\bibnamefont
  {\ifmmode \check{Z}\else \v{Z}\fi{}uti\ifmmode~\acute{c}\else \'{c}\fi{}}},
  \bibinfo {author} {\bibfnamefont {J.}~\bibnamefont {Fabian}}, \ and\ \bibinfo
  {author} {\bibfnamefont {S.}~\bibnamefont {Das~Sarma}},\ }\href {\doibase
  10.1103/RevModPhys.76.323} {\bibfield  {journal} {\bibinfo  {journal} {Rev.
  Mod. Phys.}\ }\textbf {\bibinfo {volume} {76}},\ \bibinfo {pages} {323}
  (\bibinfo {year} {2004})}\BibitemShut {NoStop}%
\bibitem [{\citenamefont {Khandekar}\ and\ \citenamefont
  {Jacob}(2019{\natexlab{a}})}]{khandekar2019spin}%
  \BibitemOpen
  \bibfield  {author} {\bibinfo {author} {\bibfnamefont {C.}~\bibnamefont
  {Khandekar}}\ and\ \bibinfo {author} {\bibfnamefont {Z.}~\bibnamefont
  {Jacob}},\ }\href
  {https://iopscience.iop.org/article/10.1088/1367-2630/ab494d} {\bibfield
  {journal} {\bibinfo  {journal} {New J. Phys.}\ }\textbf {\bibinfo {volume}
  {21}},\ \bibinfo {pages} {103030} (\bibinfo {year}
  {2019}{\natexlab{a}})}\BibitemShut {NoStop}%
\bibitem [{\citenamefont {Van~Mechelen}\ and\ \citenamefont
  {Jacob}(2016)}]{van2016universal}%
  \BibitemOpen
  \bibfield  {author} {\bibinfo {author} {\bibfnamefont {T.}~\bibnamefont
  {Van~Mechelen}}\ and\ \bibinfo {author} {\bibfnamefont {Z.}~\bibnamefont
  {Jacob}},\ }\href
  {https://www.osapublishing.org/optica/abstract.cfm?uri=optica-3-2-118}
  {\bibfield  {journal} {\bibinfo  {journal} {Optica}\ }\textbf {\bibinfo
  {volume} {3}},\ \bibinfo {pages} {118} (\bibinfo {year} {2016})}\BibitemShut
  {NoStop}%
\bibitem [{\citenamefont {Yariv}\ and\ \citenamefont
  {Yeh}(1984)}]{yariv1984optical}%
  \BibitemOpen
  \bibfield  {author} {\bibinfo {author} {\bibfnamefont {A.}~\bibnamefont
  {Yariv}}\ and\ \bibinfo {author} {\bibfnamefont {P.}~\bibnamefont {Yeh}},\
  }\href {https://books.google.com/books?id=jjzxAAAAMAAJ} {\emph {\bibinfo
  {title} {Optical waves in crystals: propagation and control of laser
  radiation}}},\ Wiley series in pure and applied optics\ (\bibinfo
  {publisher} {Wiley},\ \bibinfo {year} {1984})\BibitemShut {NoStop}%
\bibitem [{\citenamefont {Wang}\ \emph {et~al.}(2016)\citenamefont {Wang},
  \citenamefont {Cheng}, \citenamefont {Winsor},\ and\ \citenamefont
  {Liu}}]{wang2016optical}%
  \BibitemOpen
  \bibfield  {author} {\bibinfo {author} {\bibfnamefont {Z.}~\bibnamefont
  {Wang}}, \bibinfo {author} {\bibfnamefont {F.}~\bibnamefont {Cheng}},
  \bibinfo {author} {\bibfnamefont {T.}~\bibnamefont {Winsor}}, \ and\ \bibinfo
  {author} {\bibfnamefont {Y.}~\bibnamefont {Liu}},\ }\href
  {https://iopscience.iop.org/article/10.1088/0957-4484/27/41/412001/pdf}
  {\bibfield  {journal} {\bibinfo  {journal} {Nanotechnology}\ }\textbf
  {\bibinfo {volume} {27}},\ \bibinfo {pages} {412001} (\bibinfo {year}
  {2016})}\BibitemShut {NoStop}%
\bibitem [{\citenamefont {Asadchy}\ \emph {et~al.}(2018)\citenamefont
  {Asadchy}, \citenamefont {D{\'\i}az-Rubio},\ and\ \citenamefont
  {Tretyakov}}]{asadchy2018bianisotropic}%
  \BibitemOpen
  \bibfield  {author} {\bibinfo {author} {\bibfnamefont {V.}~\bibnamefont
  {Asadchy}}, \bibinfo {author} {\bibfnamefont {A.}~\bibnamefont
  {D{\'\i}az-Rubio}}, \ and\ \bibinfo {author} {\bibfnamefont {S.}~\bibnamefont
  {Tretyakov}},\ }\href
  {https://www.degruyter.com/view/j/nanoph.2018.7.issue-6/nanoph-2017-0132/nanoph-2017-0132.xml}
  {\bibfield  {journal} {\bibinfo  {journal} {Nanophotonics}\ }\textbf
  {\bibinfo {volume} {7}},\ \bibinfo {pages} {1069} (\bibinfo {year}
  {2018})}\BibitemShut {NoStop}%
\bibitem [{\citenamefont {Ishimaru}(2017)}]{ishimaru2017electromagnetic}%
  \BibitemOpen
  \bibfield  {author} {\bibinfo {author} {\bibfnamefont {A.}~\bibnamefont
  {Ishimaru}},\ }\href {https://books.google.com/books?id=NvUtDwAAQBAJ} {\emph
  {\bibinfo {title} {Electromagnetic Wave Propagation, Radiation, and
  Scattering: From Fundamentals to Applications}}},\ IEEE Press Series on
  Electromagnetic Wave Theory\ (\bibinfo  {publisher} {Wiley},\ \bibinfo {year}
  {2017})\BibitemShut {NoStop}%
\bibitem [{\citenamefont {Armelles}\ \emph {et~al.}(2013)\citenamefont
  {Armelles}, \citenamefont {Cebollada}, \citenamefont
  {Garc{\'\i}a-Mart{\'\i}n},\ and\ \citenamefont
  {Gonz{\'a}lez}}]{armelles2013magnetoplasmonics}%
  \BibitemOpen
  \bibfield  {author} {\bibinfo {author} {\bibfnamefont {G.}~\bibnamefont
  {Armelles}}, \bibinfo {author} {\bibfnamefont {A.}~\bibnamefont {Cebollada}},
  \bibinfo {author} {\bibfnamefont {A.}~\bibnamefont
  {Garc{\'\i}a-Mart{\'\i}n}}, \ and\ \bibinfo {author} {\bibfnamefont
  {M.}~\bibnamefont {Gonz{\'a}lez}},\ }\href
  {https://onlinelibrary.wiley.com/doi/epdf/10.1002/adom.201200011} {\bibfield
  {journal} {\bibinfo  {journal} {Adv. Opt. Mater.}\ }\textbf {\bibinfo
  {volume} {1}},\ \bibinfo {pages} {10} (\bibinfo {year} {2013})}\BibitemShut
  {NoStop}%
\bibitem [{\citenamefont {Sengupta}\ \emph {et~al.}(2020)\citenamefont
  {Sengupta}, \citenamefont {Khandekar}, \citenamefont {Van~Mechelen},
  \citenamefont {Rahman},\ and\ \citenamefont {Jacob}}]{electron2019sengupta}%
  \BibitemOpen
  \bibfield  {author} {\bibinfo {author} {\bibfnamefont {P.}~\bibnamefont
  {Sengupta}}, \bibinfo {author} {\bibfnamefont {C.}~\bibnamefont {Khandekar}},
  \bibinfo {author} {\bibfnamefont {T.}~\bibnamefont {Van~Mechelen}}, \bibinfo
  {author} {\bibfnamefont {R.}~\bibnamefont {Rahman}}, \ and\ \bibinfo {author}
  {\bibfnamefont {Z.}~\bibnamefont {Jacob}},\ }\href {\doibase
  10.1103/PhysRevB.101.035412} {\bibfield  {journal} {\bibinfo  {journal}
  {Phys. Rev. B}\ }\textbf {\bibinfo {volume} {101}},\ \bibinfo {pages}
  {035412} (\bibinfo {year} {2020})}\BibitemShut {NoStop}%
\bibitem [{\citenamefont {Kotov}\ and\ \citenamefont
  {Lozovik}(2018)}]{kotov2018giant}%
  \BibitemOpen
  \bibfield  {author} {\bibinfo {author} {\bibfnamefont {O.~V.}\ \bibnamefont
  {Kotov}}\ and\ \bibinfo {author} {\bibfnamefont {Y.~E.}\ \bibnamefont
  {Lozovik}},\ }\href {\doibase 10.1103/PhysRevB.98.195446} {\bibfield
  {journal} {\bibinfo  {journal} {Phys. Rev. B}\ }\textbf {\bibinfo {volume}
  {98}},\ \bibinfo {pages} {195446} (\bibinfo {year} {2018})}\BibitemShut
  {NoStop}%
\bibitem [{\citenamefont {Rodrigue}(1988)}]{rodrigue1988generation}%
  \BibitemOpen
  \bibfield  {author} {\bibinfo {author} {\bibfnamefont {G.}~\bibnamefont
  {Rodrigue}},\ }\href {https://ieeexplore.ieee.org/abstract/document/4389}
  {\bibfield  {journal} {\bibinfo  {journal} {Proc. of the IEEE}\ }\textbf
  {\bibinfo {volume} {76}},\ \bibinfo {pages} {121} (\bibinfo {year}
  {1988})}\BibitemShut {NoStop}%
\bibitem [{\citenamefont {Pyatakov}\ and\ \citenamefont
  {Zvezdin}(2012)}]{pyatakov2012magnetoelectric}%
  \BibitemOpen
  \bibfield  {author} {\bibinfo {author} {\bibfnamefont {A.}~\bibnamefont
  {Pyatakov}}\ and\ \bibinfo {author} {\bibfnamefont {A.}~\bibnamefont
  {Zvezdin}},\ }\href
  {https://iopscience.iop.org/article/10.3367/UFNe.0182.201206b.0593/meta}
  {\bibfield  {journal} {\bibinfo  {journal} {Phys.-Uspekhi}\ }\textbf
  {\bibinfo {volume} {55}},\ \bibinfo {pages} {557} (\bibinfo {year}
  {2012})}\BibitemShut {NoStop}%
\bibitem [{\citenamefont {LaForge}\ \emph {et~al.}(2010)\citenamefont
  {LaForge}, \citenamefont {Frenzel}, \citenamefont {Pursley}, \citenamefont
  {Lin}, \citenamefont {Liu}, \citenamefont {Shi},\ and\ \citenamefont
  {Basov}}]{laforge2010optical}%
  \BibitemOpen
  \bibfield  {author} {\bibinfo {author} {\bibfnamefont {A.}~\bibnamefont
  {LaForge}}, \bibinfo {author} {\bibfnamefont {A.}~\bibnamefont {Frenzel}},
  \bibinfo {author} {\bibfnamefont {B.}~\bibnamefont {Pursley}}, \bibinfo
  {author} {\bibfnamefont {T.}~\bibnamefont {Lin}}, \bibinfo {author}
  {\bibfnamefont {X.}~\bibnamefont {Liu}}, \bibinfo {author} {\bibfnamefont
  {J.}~\bibnamefont {Shi}}, \ and\ \bibinfo {author} {\bibfnamefont
  {D.}~\bibnamefont {Basov}},\ }\href
  {https://journals.aps.org/prb/abstract/10.1103/PhysRevB.81.125120} {\bibfield
   {journal} {\bibinfo  {journal} {Phys. Rev. B}\ }\textbf {\bibinfo {volume}
  {81}},\ \bibinfo {pages} {125120} (\bibinfo {year} {2010})}\BibitemShut
  {NoStop}%
\bibitem [{\citenamefont {Hu}\ \emph {et~al.}(2017)\citenamefont {Hu},
  \citenamefont {Duan}, \citenamefont {Nan},\ and\ \citenamefont
  {Chen}}]{hu2017understanding}%
  \BibitemOpen
  \bibfield  {author} {\bibinfo {author} {\bibfnamefont {J.-M.}\ \bibnamefont
  {Hu}}, \bibinfo {author} {\bibfnamefont {C.-G.}\ \bibnamefont {Duan}},
  \bibinfo {author} {\bibfnamefont {C.-W.}\ \bibnamefont {Nan}}, \ and\
  \bibinfo {author} {\bibfnamefont {L.-Q.}\ \bibnamefont {Chen}},\ }\href
  {https://www.nature.com/articles/s41524-017-0020-4} {\bibfield  {journal}
  {\bibinfo  {journal} {NPJ Comput. Mater.}\ }\textbf {\bibinfo {volume} {3}},\
  \bibinfo {pages} {1} (\bibinfo {year} {2017})}\BibitemShut {NoStop}%
\bibitem [{\citenamefont {Green}(2012)}]{green2012time}%
  \BibitemOpen
  \bibfield  {author} {\bibinfo {author} {\bibfnamefont {M.}~\bibnamefont
  {Green}},\ }\href {https://pubs.acs.org/doi/abs/10.1021/nl3034784} {\bibfield
   {journal} {\bibinfo  {journal} {Nano Lett.}\ }\textbf {\bibinfo {volume}
  {12}},\ \bibinfo {pages} {5985} (\bibinfo {year} {2012})}\BibitemShut
  {NoStop}%
\bibitem [{\citenamefont {Khandekar}\ and\ \citenamefont
  {Jacob}(2019{\natexlab{b}})}]{khandekar2019circularly}%
  \BibitemOpen
  \bibfield  {author} {\bibinfo {author} {\bibfnamefont {C.}~\bibnamefont
  {Khandekar}}\ and\ \bibinfo {author} {\bibfnamefont {Z.}~\bibnamefont
  {Jacob}},\ }\href {\doibase 10.1103/PhysRevApplied.12.014053} {\bibfield
  {journal} {\bibinfo  {journal} {Phys. Rev. Appl.}\ }\textbf {\bibinfo
  {volume} {12}},\ \bibinfo {pages} {014053} (\bibinfo {year}
  {2019}{\natexlab{b}})}\BibitemShut {NoStop}%
\bibitem [{\citenamefont {Maghrebi}\ \emph {et~al.}(2019)\citenamefont
  {Maghrebi}, \citenamefont {Gorshkov},\ and\ \citenamefont
  {Sau}}]{maghrebi2019fluctuation}%
  \BibitemOpen
  \bibfield  {author} {\bibinfo {author} {\bibfnamefont {M.~F.}\ \bibnamefont
  {Maghrebi}}, \bibinfo {author} {\bibfnamefont {A.~V.}\ \bibnamefont
  {Gorshkov}}, \ and\ \bibinfo {author} {\bibfnamefont {J.~D.}\ \bibnamefont
  {Sau}},\ }\href {\doibase 10.1103/PhysRevLett.123.055901} {\bibfield
  {journal} {\bibinfo  {journal} {Phys. Rev. Lett.}\ }\textbf {\bibinfo
  {volume} {123}},\ \bibinfo {pages} {055901} (\bibinfo {year}
  {2019})}\BibitemShut {NoStop}%
\bibitem [{\citenamefont {Reid}\ \emph {et~al.}(2017)\citenamefont {Reid},
  \citenamefont {Miller}, \citenamefont {Polimeridis}, \citenamefont
  {Rodriguez}, \citenamefont {Tomlinson},\ and\ \citenamefont
  {Johnson}}]{reid2017photon}%
  \BibitemOpen
  \bibfield  {author} {\bibinfo {author} {\bibfnamefont {M.}~\bibnamefont
  {Reid}}, \bibinfo {author} {\bibfnamefont {O.}~\bibnamefont {Miller}},
  \bibinfo {author} {\bibfnamefont {A.}~\bibnamefont {Polimeridis}}, \bibinfo
  {author} {\bibfnamefont {A.}~\bibnamefont {Rodriguez}}, \bibinfo {author}
  {\bibfnamefont {E.}~\bibnamefont {Tomlinson}}, \ and\ \bibinfo {author}
  {\bibfnamefont {S.}~\bibnamefont {Johnson}},\ }\href@noop {} {\bibfield
  {journal} {\bibinfo  {journal} {arXiv:1708.01985}\ } (\bibinfo {year}
  {2017})}\BibitemShut {NoStop}%
\bibitem [{\citenamefont {Feigel}(2004)}]{feigel2004quantum}%
  \BibitemOpen
  \bibfield  {author} {\bibinfo {author} {\bibfnamefont {A.}~\bibnamefont
  {Feigel}},\ }\href
  {https://journals.aps.org/prl/abstract/10.1103/PhysRevLett.92.020404}
  {\bibfield  {journal} {\bibinfo  {journal} {Phys. Rev. Lett.}\ }\textbf
  {\bibinfo {volume} {92}},\ \bibinfo {pages} {020404} (\bibinfo {year}
  {2004})}\BibitemShut {NoStop}%
\bibitem [{\citenamefont {Van~Tiggelen}\ \emph {et~al.}(2006)\citenamefont
  {Van~Tiggelen}, \citenamefont {Rikken},\ and\ \citenamefont
  {Krsti{\'c}}}]{van2006momentum}%
  \BibitemOpen
  \bibfield  {author} {\bibinfo {author} {\bibfnamefont {B.}~\bibnamefont
  {Van~Tiggelen}}, \bibinfo {author} {\bibfnamefont {G.}~\bibnamefont
  {Rikken}}, \ and\ \bibinfo {author} {\bibfnamefont {V.}~\bibnamefont
  {Krsti{\'c}}},\ }\href
  {https://journals.aps.org/prl/abstract/10.1103/PhysRevLett.96.130402}
  {\bibfield  {journal} {\bibinfo  {journal} {Phys. Rev. Lett.}\ }\textbf
  {\bibinfo {volume} {96}},\ \bibinfo {pages} {130402} (\bibinfo {year}
  {2006})}\BibitemShut {NoStop}%
\bibitem [{\citenamefont {Birkeland}\ and\ \citenamefont
  {Brevik}(2007)}]{ole2007extraction}%
  \BibitemOpen
  \bibfield  {author} {\bibinfo {author} {\bibfnamefont {O.~J.}\ \bibnamefont
  {Birkeland}}\ and\ \bibinfo {author} {\bibfnamefont {I.}~\bibnamefont
  {Brevik}},\ }\href {\doibase 10.1103/PhysRevE.76.066605} {\bibfield
  {journal} {\bibinfo  {journal} {Phys. Rev. E}\ }\textbf {\bibinfo {volume}
  {76}},\ \bibinfo {pages} {066605} (\bibinfo {year} {2007})}\BibitemShut
  {NoStop}%
\bibitem [{\citenamefont {Kushwaha}(2001)}]{kushwaha2001plasmons}%
  \BibitemOpen
  \bibfield  {author} {\bibinfo {author} {\bibfnamefont {M.}~\bibnamefont
  {Kushwaha}},\ }\href
  {https://www.sciencedirect.com/science/article/pii/S0167572900000078}
  {\bibfield  {journal} {\bibinfo  {journal} {Surf. Sci. Rep.}\ }\textbf
  {\bibinfo {volume} {41}},\ \bibinfo {pages} {1} (\bibinfo {year}
  {2001})}\BibitemShut {NoStop}%
\bibitem [{\citenamefont {Palik}\ \emph {et~al.}(1976)\citenamefont {Palik},
  \citenamefont {Kaplan}, \citenamefont {Gammon}, \citenamefont {Kaplan},
  \citenamefont {Wallis},\ and\ \citenamefont {Quinn}}]{palik1976coupled}%
  \BibitemOpen
  \bibfield  {author} {\bibinfo {author} {\bibfnamefont {E.}~\bibnamefont
  {Palik}}, \bibinfo {author} {\bibfnamefont {R.}~\bibnamefont {Kaplan}},
  \bibinfo {author} {\bibfnamefont {R.}~\bibnamefont {Gammon}}, \bibinfo
  {author} {\bibfnamefont {H.}~\bibnamefont {Kaplan}}, \bibinfo {author}
  {\bibfnamefont {R.}~\bibnamefont {Wallis}}, \ and\ \bibinfo {author}
  {\bibfnamefont {J.}~\bibnamefont {Quinn}},\ }\href
  {https://journals.aps.org/prb/abstract/10.1103/PhysRevB.13.2497} {\bibfield
  {journal} {\bibinfo  {journal} {Phys. Rev. B}\ }\textbf {\bibinfo {volume}
  {13}},\ \bibinfo {pages} {2497} (\bibinfo {year} {1976})}\BibitemShut
  {NoStop}%
\bibitem [{\citenamefont {Chochol}\ \emph {et~al.}(2016)\citenamefont
  {Chochol}, \citenamefont {Postava}, \citenamefont {{\v{C}}ada}, \citenamefont
  {Vanwolleghem}, \citenamefont {Halaga{\v{c}}ka}, \citenamefont {Lampin},\
  and\ \citenamefont {Pi{\v{s}}tora}}]{chochol2016magneto}%
  \BibitemOpen
  \bibfield  {author} {\bibinfo {author} {\bibfnamefont {J.}~\bibnamefont
  {Chochol}}, \bibinfo {author} {\bibfnamefont {K.}~\bibnamefont {Postava}},
  \bibinfo {author} {\bibfnamefont {M.}~\bibnamefont {{\v{C}}ada}}, \bibinfo
  {author} {\bibfnamefont {M.}~\bibnamefont {Vanwolleghem}}, \bibinfo {author}
  {\bibfnamefont {L.}~\bibnamefont {Halaga{\v{c}}ka}}, \bibinfo {author}
  {\bibfnamefont {J.-F.}\ \bibnamefont {Lampin}}, \ and\ \bibinfo {author}
  {\bibfnamefont {J.}~\bibnamefont {Pi{\v{s}}tora}},\ }\href
  {https://aip.scitation.org/doi/full/10.1063/1.4968178} {\bibfield  {journal}
  {\bibinfo  {journal} {AIP Adv.}\ }\textbf {\bibinfo {volume} {6}},\ \bibinfo
  {pages} {115021} (\bibinfo {year} {2016})}\BibitemShut {NoStop}%
\bibitem [{\citenamefont {Silveirinha}\ and\ \citenamefont
  {Maslovski}(2010)}]{silveirinha2010comment}%
  \BibitemOpen
  \bibfield  {author} {\bibinfo {author} {\bibfnamefont {M.~G.}\ \bibnamefont
  {Silveirinha}}\ and\ \bibinfo {author} {\bibfnamefont {S.~I.}\ \bibnamefont
  {Maslovski}},\ }\href {\doibase 10.1103/PhysRevLett.105.189301} {\bibfield
  {journal} {\bibinfo  {journal} {Phys. Rev. Lett.}\ }\textbf {\bibinfo
  {volume} {105}},\ \bibinfo {pages} {189301} (\bibinfo {year}
  {2010})}\BibitemShut {NoStop}%
\bibitem [{\citenamefont {Greffet}\ and\ \citenamefont
  {Nieto-Vesperinas}(1998)}]{greffet1998field}%
  \BibitemOpen
  \bibfield  {author} {\bibinfo {author} {\bibfnamefont {J.-J.}\ \bibnamefont
  {Greffet}}\ and\ \bibinfo {author} {\bibfnamefont {M.}~\bibnamefont
  {Nieto-Vesperinas}},\ }\href
  {https://www.osapublishing.org/josaa/abstract.cfm?uri=josaa-15-10-2735}
  {\bibfield  {journal} {\bibinfo  {journal} {J.Opt.Soc.Am.A}\ }\textbf
  {\bibinfo {volume} {15}},\ \bibinfo {pages} {2735} (\bibinfo {year}
  {1998})}\BibitemShut {NoStop}%
\bibitem [{\citenamefont {Ott}\ \emph {et~al.}(2018)\citenamefont {Ott},
  \citenamefont {Ben-Abdallah},\ and\ \citenamefont {Biehs}}]{ott2018circular}%
  \BibitemOpen
  \bibfield  {author} {\bibinfo {author} {\bibfnamefont {A.}~\bibnamefont
  {Ott}}, \bibinfo {author} {\bibfnamefont {P.}~\bibnamefont {Ben-Abdallah}}, \
  and\ \bibinfo {author} {\bibfnamefont {S.-A.}\ \bibnamefont {Biehs}},\ }\href
  {\doibase 10.1103/PhysRevB.97.205414} {\bibfield  {journal} {\bibinfo
  {journal} {Phys. Rev. B}\ }\textbf {\bibinfo {volume} {97}},\ \bibinfo
  {pages} {205414} (\bibinfo {year} {2018})}\BibitemShut {NoStop}%
\bibitem [{\citenamefont {Barnett}\ \emph {et~al.}(2016)\citenamefont
  {Barnett}, \citenamefont {Allen}, \citenamefont {Cameron}, \citenamefont
  {Gilson}, \citenamefont {Padgett}, \citenamefont {Speirits},\ and\
  \citenamefont {Yao}}]{barnett2016natures}%
  \BibitemOpen
  \bibfield  {author} {\bibinfo {author} {\bibfnamefont {S.}~\bibnamefont
  {Barnett}}, \bibinfo {author} {\bibfnamefont {L.}~\bibnamefont {Allen}},
  \bibinfo {author} {\bibfnamefont {R.}~\bibnamefont {Cameron}}, \bibinfo
  {author} {\bibfnamefont {C.}~\bibnamefont {Gilson}}, \bibinfo {author}
  {\bibfnamefont {M.}~\bibnamefont {Padgett}}, \bibinfo {author} {\bibfnamefont
  {F.}~\bibnamefont {Speirits}}, \ and\ \bibinfo {author} {\bibfnamefont
  {A.}~\bibnamefont {Yao}},\ }\href@noop {} {\bibfield  {journal} {\bibinfo
  {journal} {J. Opt.}\ }\textbf {\bibinfo {volume} {18}},\ \bibinfo {pages}
  {064004} (\bibinfo {year} {2016})}\BibitemShut {NoStop}%
\bibitem [{\citenamefont {Barnett}\ \emph {et~al.}(2017)\citenamefont
  {Barnett}, \citenamefont {Babiker},\ and\ \citenamefont
  {Padgett}}]{barnett2017optical}%
  \BibitemOpen
  \bibfield  {author} {\bibinfo {author} {\bibfnamefont {S.}~\bibnamefont
  {Barnett}}, \bibinfo {author} {\bibfnamefont {M.}~\bibnamefont {Babiker}}, \
  and\ \bibinfo {author} {\bibfnamefont {M.}~\bibnamefont {Padgett}},\
  }\href@noop {} {\enquote {\bibinfo {title} {Optical orbital angular
  momentum},}\ } (\bibinfo {year} {2017})\BibitemShut {NoStop}%
\bibitem [{\citenamefont {Graglia}\ \emph {et~al.}(1991)\citenamefont
  {Graglia}, \citenamefont {Uslenghi},\ and\ \citenamefont
  {Zich}}]{graglia1991reflection}%
  \BibitemOpen
  \bibfield  {author} {\bibinfo {author} {\bibfnamefont {R.}~\bibnamefont
  {Graglia}}, \bibinfo {author} {\bibfnamefont {P.}~\bibnamefont {Uslenghi}}, \
  and\ \bibinfo {author} {\bibfnamefont {R.}~\bibnamefont {Zich}},\ }\href
  {https://www.tandfonline.com/doi/abs/10.1080/02726349108908273} {\bibfield
  {journal} {\bibinfo  {journal} {Electromagnetics}\ }\textbf {\bibinfo
  {volume} {11}},\ \bibinfo {pages} {193} (\bibinfo {year} {1991})}\BibitemShut
  {NoStop}%
\bibitem [{\citenamefont {Tretyakov}\ and\ \citenamefont
  {Sochava}(1994)}]{tretyakov1994reflection}%
  \BibitemOpen
  \bibfield  {author} {\bibinfo {author} {\bibfnamefont {S.~A.}\ \bibnamefont
  {Tretyakov}}\ and\ \bibinfo {author} {\bibfnamefont {A.}~\bibnamefont
  {Sochava}},\ }\href
  {https://link.springer.com/content/pdf/10.1007/BF02096579.pdf} {\bibfield
  {journal} {\bibinfo  {journal} {Int. J. Infra. Millim. Waves}\ }\textbf
  {\bibinfo {volume} {15}},\ \bibinfo {pages} {829} (\bibinfo {year}
  {1994})}\BibitemShut {NoStop}%
\bibitem [{\citenamefont {Mueller}(1971)}]{mueller1971reflection}%
  \BibitemOpen
  \bibfield  {author} {\bibinfo {author} {\bibfnamefont {R.}~\bibnamefont
  {Mueller}},\ }\href {https://aip.scitation.org/doi/abs/10.1063/1.1660535}
  {\bibfield  {journal} {\bibinfo  {journal} {J. Appl. Phys.}\ }\textbf
  {\bibinfo {volume} {42}},\ \bibinfo {pages} {2264} (\bibinfo {year}
  {1971})}\BibitemShut {NoStop}%
\bibitem [{Note1()}]{Note1}%
  \BibitemOpen
  \bibinfo {note} {This software is available on GitHub \protect \url
  {https://github.com/chinmayCK/Fresnel} under MIT license}\BibitemShut
  {NoStop}%
\bibitem [{\citenamefont {Bimonte}\ and\ \citenamefont
  {Santamato}(2007)}]{bimonte2007general}%
  \BibitemOpen
  \bibfield  {author} {\bibinfo {author} {\bibfnamefont {G.}~\bibnamefont
  {Bimonte}}\ and\ \bibinfo {author} {\bibfnamefont {E.}~\bibnamefont
  {Santamato}},\ }\href {\doibase 10.1103/PhysRevA.76.013810} {\bibfield
  {journal} {\bibinfo  {journal} {Phys. Rev. A}\ }\textbf {\bibinfo {volume}
  {76}},\ \bibinfo {pages} {013810} (\bibinfo {year} {2007})}\BibitemShut
  {NoStop}%
\bibitem [{\citenamefont {Padilla}(2007)}]{padilla2007group}%
  \BibitemOpen
  \bibfield  {author} {\bibinfo {author} {\bibfnamefont {W.}~\bibnamefont
  {Padilla}},\ }\href
  {https://www.osapublishing.org/oe/abstract.cfm?uri=oe-15-4-1639} {\bibfield
  {journal} {\bibinfo  {journal} {Opt. Exp.}\ }\textbf {\bibinfo {volume}
  {15}},\ \bibinfo {pages} {1639} (\bibinfo {year} {2007})}\BibitemShut
  {NoStop}%
\bibitem [{\citenamefont {Li}\ and\ \citenamefont
  {Fan}(2018)}]{li2018nanophotonic}%
  \BibitemOpen
  \bibfield  {author} {\bibinfo {author} {\bibfnamefont {W.}~\bibnamefont
  {Li}}\ and\ \bibinfo {author} {\bibfnamefont {S.}~\bibnamefont {Fan}},\
  }\href {https://www.osapublishing.org/oe/abstract.cfm?uri=oe-26-12-15995}
  {\bibfield  {journal} {\bibinfo  {journal} {Opt. Exp.}\ }\textbf {\bibinfo
  {volume} {26}},\ \bibinfo {pages} {15995} (\bibinfo {year}
  {2018})}\BibitemShut {NoStop}%
\end{thebibliography}%


\begin{thebibliography}{5}%
\makeatletter
\providecommand \@ifxundefined [1]{%
 \@ifx{#1\undefined}
}%
\providecommand \@ifnum [1]{%
 \ifnum #1\expandafter \@firstoftwo
 \else \expandafter \@secondoftwo
 \fi
}%
\providecommand \@ifx [1]{%
 \ifx #1\expandafter \@firstoftwo
 \else \expandafter \@secondoftwo
 \fi
}%
\providecommand \natexlab [1]{#1}%
\providecommand \enquote  [1]{``#1''}%
\providecommand \bibnamefont  [1]{#1}%
\providecommand \bibfnamefont [1]{#1}%
\providecommand \citenamefont [1]{#1}%
\providecommand \href@noop [0]{\@secondoftwo}%
\providecommand \href [0]{\begingroup \@sanitize@url \@href}%
\providecommand \@href[1]{\@@startlink{#1}\@@href}%
\providecommand \@@href[1]{\endgroup#1\@@endlink}%
\providecommand \@sanitize@url [0]{\catcode `\\12\catcode `\$12\catcode
  `\&12\catcode `\#12\catcode `\^12\catcode `\_12\catcode `\%12\relax}%
\providecommand \@@startlink[1]{}%
\providecommand \@@endlink[0]{}%
\providecommand \url  [0]{\begingroup\@sanitize@url \@url }%
\providecommand \@url [1]{\endgroup\@href {#1}{\urlprefix }}%
\providecommand \urlprefix  [0]{URL }%
\providecommand \Eprint [0]{\href }%
\providecommand \doibase [0]{http://dx.doi.org/}%
\providecommand \selectlanguage [0]{\@gobble}%
\providecommand \bibinfo  [0]{\@secondoftwo}%
\providecommand \bibfield  [0]{\@secondoftwo}%
\providecommand \translation [1]{[#1]}%
\providecommand \BibitemOpen [0]{}%
\providecommand \bibitemStop [0]{}%
\providecommand \bibitemNoStop [0]{.\EOS\space}%
\providecommand \EOS [0]{\spacefactor3000\relax}%
\providecommand \BibitemShut  [1]{\csname bibitem#1\endcsname}%
\let\auto@bib@innerbib\@empty
\bibitem [{\citenamefont {Kr{\"u}ger}\ \emph {et~al.}(2012)\citenamefont
  {Kr{\"u}ger}, \citenamefont {Bimonte}, \citenamefont {Emig},\ and\
  \citenamefont {Kardar}}]{kruger2012trace}%
  \BibitemOpen
  \bibfield  {author} {\bibinfo {author} {\bibfnamefont {M.}~\bibnamefont
  {Kr{\"u}ger}}, \bibinfo {author} {\bibfnamefont {G.}~\bibnamefont {Bimonte}},
  \bibinfo {author} {\bibfnamefont {T.}~\bibnamefont {Emig}}, \ and\ \bibinfo
  {author} {\bibfnamefont {M.}~\bibnamefont {Kardar}},\ }\bibfield  {title}
  {\enquote {\bibinfo {title} {Trace formulas for nonequilibrium casimir
  interactions, heat radiation, and heat transfer for arbitrary objects},}\
  }\href {https://journals.aps.org/prb/abstract/10.1103/PhysRevB.86.115423}
  {\bibfield  {journal} {\bibinfo  {journal} {Phys. Rev. B}\ }\textbf {\bibinfo
  {volume} {86}},\ \bibinfo {pages} {115423} (\bibinfo {year}
  {2012})}\BibitemShut {NoStop}%
\bibitem [{\citenamefont {Maghrebi}\ \emph {et~al.}(2012)\citenamefont
  {Maghrebi}, \citenamefont {Jaffe},\ and\ \citenamefont
  {Kardar}}]{maghrebi2012spontaneous}%
  \BibitemOpen
  \bibfield  {author} {\bibinfo {author} {\bibfnamefont {M.~F.}\ \bibnamefont
  {Maghrebi}}, \bibinfo {author} {\bibfnamefont {R.~L.}\ \bibnamefont {Jaffe}},
  \ and\ \bibinfo {author} {\bibfnamefont {M.}~\bibnamefont {Kardar}},\
  }\bibfield  {title} {\enquote {\bibinfo {title} {Spontaneous emission by
  rotating objects: A scattering approach},}\ }\href {\doibase
  10.1103/PhysRevLett.108.230403} {\bibfield  {journal} {\bibinfo  {journal}
  {Phys. Rev. Lett.}\ }\textbf {\bibinfo {volume} {108}},\ \bibinfo {pages}
  {230403} (\bibinfo {year} {2012})}\BibitemShut {NoStop}%
\bibitem [{\citenamefont {Bimonte}(2009)}]{bimonte2009scattering}%
  \BibitemOpen
  \bibfield  {author} {\bibinfo {author} {\bibfnamefont {G.}~\bibnamefont
  {Bimonte}},\ }\bibfield  {title} {\enquote {\bibinfo {title} {Scattering
  approach to casimir forces and radiative heat transfer for nanostructured
  surfaces out of thermal equilibrium},}\ }\href {\doibase
  10.1103/PhysRevA.80.042102} {\bibfield  {journal} {\bibinfo  {journal} {Phys.
  Rev. A.}\ }\textbf {\bibinfo {volume} {80}},\ \bibinfo {pages} {042102}
  (\bibinfo {year} {2009})}\BibitemShut {NoStop}%
\bibitem [{\citenamefont {Joulain}\ \emph {et~al.}(2005)\citenamefont
  {Joulain}, \citenamefont {Mulet}, \citenamefont {Marquier}, \citenamefont
  {Carminati},\ and\ \citenamefont {Greffet}}]{joulain2005surface}%
  \BibitemOpen
  \bibfield  {author} {\bibinfo {author} {\bibfnamefont {K.}~\bibnamefont
  {Joulain}}, \bibinfo {author} {\bibfnamefont {J-P.}\ \bibnamefont {Mulet}},
  \bibinfo {author} {\bibfnamefont {F.}~\bibnamefont {Marquier}}, \bibinfo
  {author} {\bibfnamefont {R.}~\bibnamefont {Carminati}}, \ and\ \bibinfo
  {author} {\bibfnamefont {J.-J.}\ \bibnamefont {Greffet}},\ }\bibfield
  {title} {\enquote {\bibinfo {title} {Surface electromagnetic waves thermally
  excited: Radiative heat transfer, coherence properties and casimir forces
  revisited in the near field},}\ }\href@noop {} {\bibfield  {journal}
  {\bibinfo  {journal} {Surf. Sci. Rep.}\ }\textbf {\bibinfo {volume} {57}},\
  \bibinfo {pages} {59--112} (\bibinfo {year} {2005})}\BibitemShut {NoStop}%
\bibitem [{\citenamefont {Moncada-Villa}\ \emph {et~al.}(2015)\citenamefont
  {Moncada-Villa}, \citenamefont {Fern\'andez-Hurtado}, \citenamefont
  {Garc\'{\i}a-Vidal}, \citenamefont {Garc\'{\i}a-Mart\'{\i}n},\ and\
  \citenamefont {Cuevas}}]{moncada2015magnetic}%
  \BibitemOpen
  \bibfield  {author} {\bibinfo {author} {\bibfnamefont {E.}~\bibnamefont
  {Moncada-Villa}}, \bibinfo {author} {\bibfnamefont {V.}~\bibnamefont
  {Fern\'andez-Hurtado}}, \bibinfo {author} {\bibfnamefont {F.~J.}\
  \bibnamefont {Garc\'{\i}a-Vidal}}, \bibinfo {author} {\bibfnamefont
  {A.}~\bibnamefont {Garc\'{\i}a-Mart\'{\i}n}}, \ and\ \bibinfo {author}
  {\bibfnamefont {J.~C.}\ \bibnamefont {Cuevas}},\ }\bibfield  {title}
  {\enquote {\bibinfo {title} {Magnetic field control of near-field radiative
  heat transfer and the realization of highly tunable hyperbolic thermal
  emitters},}\ }\href {\doibase 10.1103/PhysRevB.92.125418} {\bibfield
  {journal} {\bibinfo  {journal} {Phys. Rev. B}\ }\textbf {\bibinfo {volume}
  {92}},\ \bibinfo {pages} {125418} (\bibinfo {year} {2015})}\BibitemShut
  {NoStop}%
\end{thebibliography}%
 
\end{document}


\title{Universal spin-resolved thermal radiation laws for
  nonreciprocal bianisotropic media: Supplementary Materials}

\author{Chinmay Khandekar} \email{ckhandek@purdue.edu}
\affiliation{Birck Nanotechnology Center, School of Electrical and
  Computer Engineering, College of Engineering, Purdue University,
  West Lafayette, Indiana 47907, USA}

\author{Farhad Khosravi} \affiliation{Department of Electrical and
  Computer Engineering, University of Alberta, Edmonton, Alberta
  T6G1H9, Canada}

\author{Zhou Li} 
\affiliation{Birck Nanotechnology Center, School of Electrical and
  Computer Engineering, College of Engineering, Purdue University,
  West Lafayette, Indiana 47907, USA}
    
\author{Zubin Jacob} \email{zjacob@purdue.edu}
\affiliation{Birck Nanotechnology Center, School of Electrical and
  Computer Engineering, College of Engineering, Purdue University,
  West Lafayette, Indiana 47907, USA}

\date{\today}

\begin{abstract}
  We show the derivation of spin-resolved emissivities using Rytov's
  fluctuational electrodynamics. We provide a universal perspective of
  spin-resolved thermal-radiation characteristics of bi-layered
  bianisotropic planar media; an extension of the same analysis for
  single-layered slabs in the main text. Finally, we derive the
  closed-form expressions of Fresnel coefficients for a semi-infinite
  half-space of gyromagnetic medium under specific conditions.
\end{abstract}

\pacs{} \maketitle

\onecolumngrid

\section{Fluctuational electrodynamic derivation of emissivity}

The following basis-independent trace formula for the thermal emission
power from an object at temperature $T$ into the external vacuum at
temperature $T_e$ in terms of its scattering operator ($S$) has been
derived previously~\cite{kruger2012trace,maghrebi2012spontaneous,
  bimonte2009scattering}:
\begin{align}
H = \int \frac{d\omega}{2\pi} [\Theta(\omega,T)-\Theta(\omega,T_e)]
\text{Tr}\{\mathcal{I} - SS^{\dagger}\}
\label{scatter}
\end{align}
Substituting the scattering matrix for the planar interface in the
basis of RCP ($\ev_{(+)}$) and LCP ($\ev_{(-)}$) plane waves, we
obtain:
\begin{align}
H_s &= \int\frac{d\omega}{2\pi} [\Theta(\omega,T)-\Theta(\omega,T_e)]
\text{Tr}\{ \int_0^{k_0=\omega/c}\frac{k_\parallel dk_{\parallel}}{(2\pi)^2}
\int_0^{2\pi} d\phi \bigg( \mathcal{I}- \begin{bmatrix} r_{(++)}
  & r_{(+-)} \\ r_{(-+)} & r_{(--)}\end{bmatrix} \begin{bmatrix}
  r^*_{(++)} & r^*_{(-+)} \\ r^*_{(+-)} & r^*_{(--)}\end{bmatrix} \bigg) \}
\end{align}
Here $H_s$ is the Poynting flux or heat radiation per unit area of the
slab in the normal ($\ev_z$) direction and $r_{(jk)}$ for $j,k=[+,-]$
denotes the amplitude of j-polarized reflected circularly polarized
(CP) wave due to unit amplitude k-polarized incident CP
wave. $R_{(jk)}=|r_{(jk)}|^2$ denotes the polarization interconversion
reflectance used in the calculation of emissivities and
absorptivities. $k_{\parallel}$ is the in-plane wavevector which is
related to the vacuum wavevector $k_0=\omega/c$ as $k_\parallel = k_0
\sin\theta$. Using $d\Omega=\sin\theta d\theta d\phi$ and assuming the
surrounding to be at $T_e=0$K, we obtain:
\begin{align}
H_s = \int d\omega \int d\Omega
[\underbrace{1-R_{(++)}(\theta,\phi)-R_{(+-)}(\theta,\phi)}_{
    \eta_{(+)}(\theta,\phi+\pi)} +
  \underbrace{1-R_{(-+)}(\theta,\phi)-R_{(--)}(\theta,\phi)}_{
    \eta_{(-)}(\theta,\phi+\pi)}] \frac{I_b(\omega,T)}{2}\cos\theta
\end{align}
Here $I_b(\omega,T)=\omega^2\Theta(\omega,T)/(4\pi^3 c^2)$ is the
blackbody radiance at temperature $T$ and
$\Theta(\omega,T)=\hbar\omega/[\text{exp}(\hbar\omega/k_B T)-1]$ is
the Planck's function. It follows that the integrand of the above
expression reproduces the expression for the emitted power per unit
area ($dA$), per unit frequency $d\omega$ within the solid angle
$d\Omega$ given by Eq.1 in manuscript. We note that this derivation is
not new and has been presented before for isotropic planar media and
in the usual basis of ($\bm{s},\bm{p}$)-polarization
states~\cite{kruger2012trace}. A derivation based on the second kind
of fluctuation-dissipation theorem (correlations of underlying
fluctuating current densities) can be found for isotropic,
dielectric/metallic materials in Ref.~\cite{joulain2005surface} and
for nonreciprocal gyroelectric or magneto-optic planar media in
Ref.~\cite{moncada2015magnetic}.

\section{Universal perspective of spin-resolved thermal radiation
  from multilayered slabs}

\begin{figure*}[t!]
  \centering\includegraphics[width=0.95\linewidth]{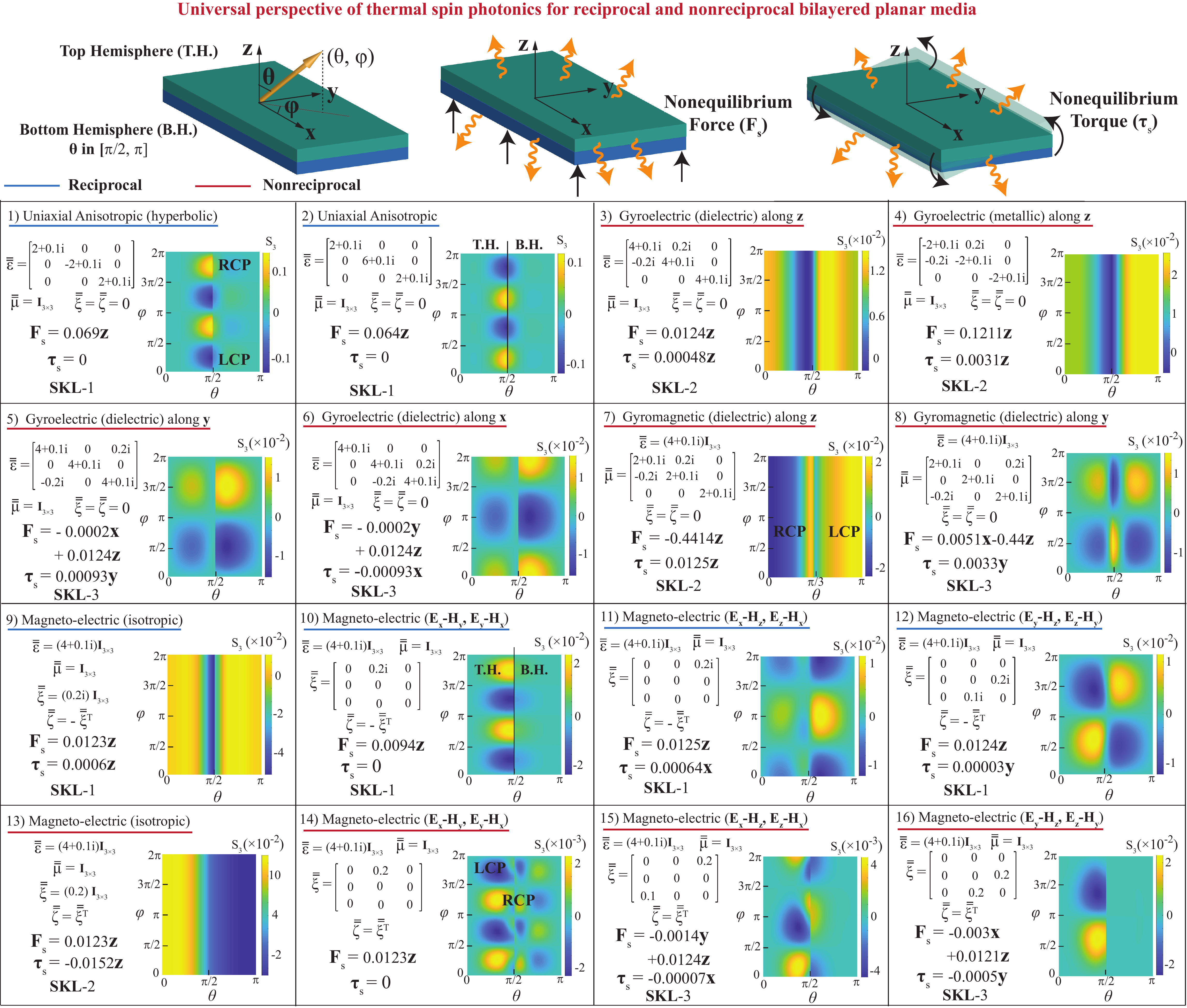}
  \caption{This figure (an extension of figure 3 in the manuscript)
    analyzes circularly polarized thermal emission and associated
    nonequilibrium force and torque for a bi-layered finite-thickness
    slab. We consider thermal emission at frequency $\omega$ from the
    slab which consists of a bianisotropic material described by
    $\epsb,\mub,\xib,\zetab$ parameters (top layer having thickness
    $d=0.5c/\omega$) on top of a trivial dielectric material of
    permittivity $\epsb=(8+0.1i)\mathcal{I}_{3\times 3}$ (bottom layer
    having thickness $d=0.4c/\omega$). The bi-layered slab at
    temperature $T_0$ emits thermal radiation into the surrounding
    vacuum on both sides at temperature $T=0$K. The contour plot in
    each representative example demonstrates the Stokes $S_3$
    parameter (Eq.\ref{stokes}) as a function of $(\theta,\phi)$ which
    are demonstrated in the top left inset. The calculated spectral
    force per unit area $\bm{F}_s/(I_{b,T_0}/2c)$ and the spectral
    torque per unit area $\bm{\tau}_s/(I_{b,T_0}/2\omega)$ arising
    from the thermal emission from the slab are provided for the given
    material parameters. The applicable spin-resolved Kirchhoff's law
    ({\bf SKL}) is also noted for each example.}
  \label{fig1s}
\end{figure*}

In the main text, we provided a universal perspective of spin-resolved
thermal radiation from finite-thickness planar slabs of many material
classes. Here we extend it to the case of multi-layered planar
slabs. In figure 3 of the manuscript, we consider spin-resolved
thermal emission at frequency $\omega$ from a planar slab of thickness
$d=0.5c/\omega$ containing a bianisotropic medium characterized by the
response parameters $\epsb,\mub,\xib,\zetab$. Here, in
figure~\ref{fig1s}, we use the exact same parameters for all
examples. But, additionally, we consider the slab to be placed on a
dielectric substrate of thickness $d=0.4c/\omega$ having permittivity
$\epsb=(8+0.1i)\mathcal{I}_{3\times 3}$. This asymmetric arrangement
(along $\bm{z}$-direciton) leads to a nonzero spectral thermal force
perpendicular to the slab which is otherwise absent for single-layered
slabs of most material classes. As explained in the main text, the
asymmetric arrangement also leads to a nonzero perpendicular thermal
torque for slabs containing isotropic magnetoelectric media (see
examples 9,13) where single-layered slabs of these materials do not
experience any torque due to symmetric spin-polarized emission in top
and bottom hemispheres.

\emph{Parallel forces and torques:} In the main text, we find that,
without using the substrate slab, the parallel nonequilibrium force is
experienced only by anisotropic nonreciprocal magnetoelectric
materials which cause coupling between perpendicular and parallel
components of $\Ev-\Hv$ fields (examples 15 and 16).
Figure~\ref{fig1s} reveals that, in a bi-layered system, gyrotropic
media with gyrotropy axis parallel to the surface (examples 5,6,8)
also experience a parallel nonequilibrium force. Furthermore, the
above anisotropic magnetoelectric materials (examples 15,16) also
experience a parallel nonequilibrium torque that is otherwise absent
in a single-layered slab. We note that the parallel forces and torques
are both absent for single-layered slabs of anisotropic reciprocal
magnetoelectric media (examples 11, 12). However, in a bi-layer
system, while the parallel force remains zero, these slabs can
experience a torque parallel to the surface. The examples in
figure~\ref{fig1s} also reveal the dependence of directionalities of
nonequilibrium force and torque on the material properties. For
example, multilayered planar slabs of gyrotropic media with gyrotropy
axis parallel to the surface (examples 5,6,8) experience a torque
parallel to the gyrotropy axis and a force which is parallel to the
surface but perpendicular to the gyrotropy axis. A multilayered planar
slab containing anisotropic nonreciprocal magnetoelectric media that
cause coupling of $\Ev_y-\Hv_z,\Ev_z-\Hv_y$ fields (example 16)
experience a force along $\bm{x}$-axis and a torque along
$\bm{y}$-axis.

Based on this universal perspective of multilayered bianisotropic
slabs which are classified according to the spin-resolved Kirchhoff's
laws ({\bf SKL}s), the following general rules can be obtained. They
are useful from the perspective of engineering thermal optomechanical
forces and torques:
\begin{itemize}
  \item Perpendicular nonequilibrium thermal force can be obtained for
    all materials classes.
  \item Perpendicular nonequilibrium thermal torque can be obtained
    for nonreciprocal media satisfying {\bf SKL-2} and isotropic
    reciprocal magnetoelectric media (Pasteur media).
  \item Parallel nonequilibrium thermal force can exist only for
    nonreciprocal materials satisfying {\bf SKL-3}.
  \item Parallel nonequilibrium thermal torque can exist for
    nonreciprocal media satisfying {\bf SKL-3} and anisotropic
    reciprocal magnetoelectric media which cause coupling between
    perpendicular and parallel components of $\Ev-\Hv$ fields.
\end{itemize}
While deriving these rules, we have omitted consideration of
multilayered slabs intermixing materials satisfying different {\bf
  SKL}s. In that case, both perpendicular and parallel components of
nonequilibrium force and torque can be engineered.

\section{Fresnel reflection coefficients for a gyromagnetic half-space}

We provide semi-analytic closed form expressions of Fresnel reflection
coefficients for a gyromagnetic medium and derive some of the
interesting relations between them noted in the main text. We consider
a semi-infinite half-space of a gyromagnetic material having
permittivity $\epsb= \epsilon \bm{I}_{3\times 3}$, and a generic
anti-symmetric permeability tensor $\mub = \mu_{ij}$ having nonzero
off-diagonal elements with the property $\mu_{ij} = - \mu_{ji}$. On
the vacuum side of the geometry, we write the incident and reflected
electric and magetic fields as the following:
\begin{subequations}
\begin{equation}
  \bm E = \bm E^i + \bm E^r, \quad \bm H = \bm H^i + \bm H^r
\end{equation}
\begin{equation}
  \bm E^i = (E_{0s} \hat{s}_- + E_{0p}\hat{p}_-)e^{i\bm k_-\cdot \bm r}
\end{equation}
\begin{equation}
  \bm E^r = \left( E_{0s} r_{ss} \hat{s}_+ + E_{0p} r_{pp} \hat{p}_+ +
  E_{0s} r_{ps} \hat{p}_+ + E_{0p} r_{sp} \hat{s}_+ \right)e^{i\bm k_+
    \cdot \bm r}
\end{equation}
\begin{equation}
  \bm H^i = \frac{1}{\eta_0}\left( - E_{0s} \hat{p}_- + E_{0p}
  \hat{s}_- \right)e^{i \bm k_- \cdot \bm r}
\end{equation}
\begin{equation}
  \bm H^r = \frac{1}{\eta_0}\left( - E_{0s} r_{ss} \hat{p}_+ + E_{0p}
  r_{pp} \hat{s}_+ + E_{0s}r_{ps} \hat{s}_+ - E_{0p} r_{sp} \hat{p}_+
  \right) e^{i \bm k_+ \cdot \bm r}
\end{equation}
\end{subequations}
where the polarization eigenstates $\hat{s}_\pm$, $\hat{p}_\pm$, and
the associated wavevectors $\hat{k}_\pm/k_0$ are:
\begin{equation}
  \hat{k}_\pm = k_0 \left( k_{\parallel} \cos\phi \hat{x} +
  k_{\parallel} \sin\phi \hat{y} \pm k_z \hat{z} \right), \quad
  \hat{s}_\pm = \sin \phi \hat{x} - \cos \phi \hat{y}, \quad
  \hat{p}_{\pm} = -\left( \pm k_z \cos\phi \hat{x} \pm k_z \sin \phi -
  k_{\parallel} \hat{z} \right)
\end{equation}
$k_\parallel$ denotes the conserved wavevector component parallel to
the surface and $\eta_0 = \sqrt{\mu_0/\epsilon_0}$. Similarly, we can
write the electric and magnetic fields inside the gyromagnetic
material as
\begin{subequations}
  \begin{equation}
    \bm E' = \bm E^t, \quad \bm H' = \bm H^t
  \end{equation}
  \begin{equation}
    \bm E^t = \left( E_{0s}t_{ss} \hat{s}'_- + E_{0p} t_{pp}
    \hat{p}'_- + E_{0s} t_{ps} \hat{p}'_- + E_{0p}t_{sp} \hat{s}'_-
    \right)e^{i\bm k'_-\cdot r}
  \end{equation}
  \begin{equation}
    \bar{\bar{\bm \mu}} \bm H^t = \frac{\sqrt{k_{\parallel}^2 +
        k_z'^2}}{\eta_0} \left[ - E_{0s} t_{ss} \hat{p}'_- + E_{0p}
      t_{pp} \hat{s}'_- + E_{0s}t_{ps} \hat{s}'_- - E_{0p} t_{sp}
      \hat{p}'_- \right]e^{i\bm k'_- \cdot \bm r }
  \end{equation}
\end{subequations}
where
\begin{equation}
  \bm{k}'_\pm = k_0 \hat{k}'_\pm = k_0 \left(k_{\parallel} \cos\phi
  \hat{x} + k_{\parallel} \sin\phi \hat{y} \pm k_z' \hat{z}\right),
  \quad \hat{s}'_\pm = \sin \phi \hat{x} - \cos \phi \hat{y}, \quad
  \hat{p}'_{\pm} = - \frac{ \pm k_z' \cos\phi \hat{x} \pm k_z' \sin
    \phi \hat{y} - k_{\parallel} \hat{z}}{\sqrt{k_{\parallel}^2 +
      k_z'^2}}
\end{equation}
Since there is only one interface, we have denoted the transmission
coefficients as $t_{jk}$ for $j,k=[s,p]$. Note that $\hat{k}'_\pm
\times \hat{p}'_\pm = \hat{s}'_\pm$. The value of $k_z'$
($\bm{z}$-component of the wavevector) inside the medium can be
obtained by solving the Maxwell's equations as described in the
methods section of the main text. From these solutions and applying
the boundary conditions for separate incidence of $s$-polarized light
and $p$-polarized light, we find the values of $r_{ss}$, $r_{sp}$,
$r_{ps}$, $r_{pp}$ for given $k_{\parallel}$ and $\phi$. For a
gyromagnetic material considered here, the following simplified
quartic equation is obtained:
\begin{equation}\label{Eq:eigen_eq}
  a \left( k_z' \right)^4 + c \left( k_z' \right)^2 + d k_z' + e = 0
\end{equation}
where
\begin{subequations}
  \begin{equation}
    a = \mu_{zz}
  \end{equation}
  \begin{equation}
    c = \mu_{zz}k_{\parallel}^2 + \mu_{xx}k_{\parallel}^2\cos^2\phi +
    \mu_{yy}k_{\parallel}^2\sin^2\phi - \epsilon\mu_{zz}\left(
    \mu_{xx} + \mu_{yy} \right) - \epsilon\mu_{yz}^2 -
    \epsilon\mu_{zx}^2
  \end{equation}
  \begin{equation}\label{Eq:d_eigen_eq}
    d = 2\epsilon\mu_{xy}k_{\parallel} \left( \mu_{yz}\cos\phi +
    \mu_{zx} \sin\phi \right)
  \end{equation}
  \begin{equation}
  \begin{split}
    e = & k_{\parallel}^4 \left(\mu_{xx}\cos^2\phi +
    \mu_{yy}\sin^2\phi\right) + \epsilon^2 \mu_{xx}\mu_{yy}\mu_{zz} -
    \epsilon\mu_{xx}\mu_{yy}k_{\parallel}^2 -
    \epsilon\mu_{zz}k_{\parallel}^2\left( \mu_{xx}\cos^2\phi +
    \mu_{yy}\sin^2\phi \right) \\ + & \epsilon^2 \mu_{xx}\mu_{yz}^2 +
    \epsilon^2 \mu_{yy}\mu_{zx}^2 + \epsilon^2 \mu_{zz}\mu_{xy}^2 -
    \epsilon\mu_{xy}^2k_{\parallel}^2 -
    \epsilon k_{\parallel}^2\left(\mu_{zx}\cos\phi +
    \mu_{yz}\sin\phi\right)^2
  \end{split}
  \end{equation}
\end{subequations}
There are four solutions where two of them correspond to waves
propagating in $-\bm{z}$ direction and the other two correspond to
waves propagating in $+\bm{z}$ direction. We choose the solutions that
correspond to decay of the fields away from the interface inside the
medium. Identifying these solutions as $k_z^{'(i)}$ with $i=1,2$, we
have two sets of $t_{ss}^{(i)}$, $t_{sp}^{(i)}$, $t_{ps}^{(i)}$, and
$t_{pp}^{(i)}$. Substituting the associated fields in Maxwell's
equations, we obtain following equations relating these coefficients
by considering $\bm{x}$ component of the fields:
\begin{equation}
  \begin{split}
      & t_{ss}^{(i)} E_{0s} \left[ \epsilon \sin\phi +
      \left(k_z^{'(i)}\right)^2 \mu_{xy}^{-1} \cos\phi -
      \mu_{yy}^{-1}\left( k_z^{'(i)}\right)^2 \sin\phi -
      \mu_{yz}^{-1}k_{\parallel} k_z^{'(i)} -
      \mu_{zx}^{-1}k_{\parallel} k_z^{'(i)} \sin\phi\cos\phi +
      \mu_{yz}^{-1}k_{\parallel} k_z^{'(i)}\sin^2\phi -
      \mu_{zz}^{-1}k_{\parallel}^2\sin\phi \right] \\ + & t_{pp}^{(i)}
    E_{0p} \sqrt{k_{\parallel}^2 + {k_z^{'(i)}}^2} \left[
      \frac{\epsilon k_z^{'(i)}\cos\phi}{\left(k_{\parallel}^2 +
        {k_z^{'(i)}}^2\right)} - \mu_{xy}^{-1}k_z^{'(i)}\sin\phi -
      \mu_{yy}^{-1}k_z^{'(i)}\cos\phi +
      \mu_{zx}^{-1}k_{\parallel}\sin^2\phi +
      \mu_{yz}^{-1}k_{\parallel} \sin\phi \cos\phi \right] \\ + &
    t_{ps}^{(i)} E_{0s} \sqrt{k_{\parallel}^2 + {k_z^{'(i)}}^2} \left[
      \frac{\epsilon k_z^{'(i)}\cos\phi}{\left(k_{\parallel}^2 +
        {k_z^{'(i)}}^2\right)} - \mu_{xy}^{-1} k_z^{'(i)}\sin\phi -
      \mu_{yy}^{-1}k_z^{'(i)}\cos\phi + \mu_{zx}^{-1}k_{\parallel}
      \sin^2\phi + \mu_{yz}^{-1}k_{\parallel}\sin\phi\cos\phi \right]
    \\ + & t_{sp}^{(i)} E_{0p}\left[ \epsilon \sin\phi +
      \mu_{xy}^{-1}\left(k_z^{'(i)}\right)^2 \cos\phi
      -\mu_{yy}^{-1}\left( k_z^{'(i)} \right)^2 \sin\phi -
      \mu_{yz}^{-1}k_{\parallel} k_z^{'(i)} -
      \mu_{zx}^{-1}k_{\parallel} k_z^{'(i)} \sin\phi \cos\phi +
      \mu_{yz}^{-1}k_{\parallel} k_z^{'(i)}\sin^2\phi -
      \mu_{zz}^{-1}k_{\parallel}^2 \sin\phi \right] \\ = & 0
  \end{split}
\end{equation}
Setting $E_{0p}=0$ for calculation of $r_{ss},r_{ps}$ below (incidence
of $s$-polarized light), we obtain:
\begin{equation}\label{Eq:A(i)_01}
\begin{split}
  A^{(i)} = & \frac{t_{ps}^{(i)}}{t_{ss}^{(i)}} \\ = & -
  \frac{\epsilon \sin\phi + \left(k_z^{'(i)}\right)^2 \mu_{xy}^{-1}
    \cos\phi - \mu_{yy}^{-1}\left( k_z^{'(i)}\right)^2 \sin\phi -
    \mu_{yz}^{-1}k_{\parallel} k_z^{'(i)} - \mu_{zx}^{-1}k_{\parallel}
    k_z^{'(i)} \sin\phi\cos\phi + \mu_{yz}^{-1}k_{\parallel}
    k_z^{'(i)}\sin^2\phi -
    \mu_{zz}^{-1}k_{\parallel}^2\sin\phi}{\sqrt{k_{\parallel}^2 +
      {k_z^{'(i)}}^2} \left[\frac{\epsilon
        k_z^{'(i)}\cos\phi}{\left(k_{\parallel}^2 +
        {k_z^{'(i)}}^2\right)} - \mu_{xy}^{-1} k_z^{'(i)}\sin\phi -
      \mu_{yy}^{-1}k_z^{'(i)}\cos\phi + \mu_{zx}^{-1}k_{\parallel}
      \sin^2\phi + \mu_{yz}^{-1}k_{\parallel}\sin\phi\cos\phi\right]}
 \end{split}
\end{equation}
and setting $E_{0s} = 0$ (incidence of $p$-polarized light), we find:
\begin{equation}
  B^{(i)} = \frac{t_{sp}^{(i)}}{t_{pp}^{i}} = \frac{1}{A^{(i)}}
\end{equation}
We also obtain $A^{(i)}$ and $B^{(i)}$ from the $\bm{y}$-component of
the fields which are as the following:
\begin{equation}\label{Eq:A(i)_02}
\begin{split}
  A^{(i)} = & \frac{t_{ps}^{(i)}}{t_{ss}^{(i)}} \\ = & -
  \frac{-\epsilon \cos\phi + \left(k_z^{'(i)}\right)^2 \mu_{xx}^{-1}
    \cos\phi + \mu_{xy}^{-1}\left( k_z^{'(i)}\right)^2 \sin\phi -
    \mu_{zx}^{-1}k_{\parallel} k_z^{'(i)} + \mu_{zx}^{-1}k_{\parallel}
    k_z^{'(i)} \cos^2\phi - \mu_{yz}^{-1}k_{\parallel}
    k_z^{'(i)}\sin\phi\cos\phi +
    \mu_{zz}^{-1}k_{\parallel}^2\cos\phi}{\sqrt{k_{\parallel}^2 +
      {k_z^{'(i)}}^2} \left[ \frac{\epsilon
        k_z^{'(i)}\sin\phi}{\left(k_{\parallel}^2 +
        {k_z^{'(i)}}^2\right)} - \mu_{xx}^{-1} k_z^{'(i)}\sin\phi +
      \mu_{xy}^{-1}k_z^{'(i)}\cos\phi - \mu_{zx}^{-1}k_{\parallel}
      \sin\phi\cos\phi - \mu_{yz}^{-1}k_{\parallel}\cos^2\phi \right]}
 \end{split} 
\end{equation}
\begin{equation}
  B^{(i)} = \frac{t_{sp}^{(i)}}{t_{pp}^{(i)}} = \frac{1}{A^{(i)}}
\end{equation}

We now write the boundary conditions for electric fields.
\begin{subequations}
  \begin{equation}
  \begin{split}
    \hat{x}: & E_{0s} \sin\phi (1 + r_{ss}) + E_{0p} k_z \cos\phi (1 -
    r_{pp}) - E_{0s}r_{ps}k_z \cos\phi + E_{0p} r_{sp} \sin\phi = \\ &
    \sum_{i=1}^{2} \left[ E_{0s} t_{ss}^{(i)} \sin\phi +
      \frac{E_{0p}t_{pp}^{(i)} k_z'^{(i)}
        \cos\phi}{\sqrt{k_{\parallel}^2 + {k_z^{'(i)}}^2}} + \frac{
        E_{0s} t_{ps}^{(i)}k_z'^{(i)} \cos\phi}{\sqrt{k_{\parallel}^2
          + {k_z^{'(i)}}^2}} + E_{0p} t_{sp}^{(i)}\sin\phi \right]
    \end{split}
  \end{equation}
  \begin{equation}
  \begin{split}
    \hat{y}: & -E_{0s}\cos\phi (1 + r_{ss}) + E_{0p}k_z \sin\phi ( 1 +
    r_{pp}) - E_{0s}r_{ps}k_z\sin\phi - E_{0p}r_{sp}\cos\phi = \\ &
    \sum_{i=1}^{2} \left[- E_{0s} t_{ss}^{(i)} \cos\phi + \frac{E_{0p}
        t_{pp}^{(i)} k_z'^{(i)}\sin\phi}{\sqrt{k_{\parallel}^2 +
          {k_z^{'(i)}}^2}} + \frac{ E_{0s} t_{ps}^{(i)}
        k_z'^{(i)}\sin\phi}{\sqrt{k_{\parallel}^2 + {k_z^{'(i)}}^2}} -
      E_{0p} t_{sp}^{(i)} \cos\phi\right].
    \end{split}
  \end{equation}
\end{subequations}
The boundary conditions for the magnetic field are:
\begin{subequations}
  \begin{equation}
  \begin{split}
    \hat{x}: E_{0s} k_z \cos\phi (1 - r_{ss}) - & E_{0p}\sin\phi (1 +
    r_{pp}) - E_{0s}r_{ps} \sin\phi - E_{0p}r_{sp} k_z\cos\phi =
    \\ \sum_{i=1}^{2} \Bigg\{& \mu^{-1}_{xx} \left[ E_{0s}
      t_{ss}^{(i)}k_z'^{(i)} \cos\phi - E_{0p}t_{pp}^{(i)}
      \sqrt{k_{\parallel}^2 + {k_z^{'(i)}}^2} \sin\phi - E_{0s}
      t_{ps}^{(i)} \sqrt{k_{\parallel}^2 + {k_z^{'(i)}}^2} \sin\phi +
      E_{0p} t_{sp}^{(i)} k_z'^{(i)} \cos\phi \right] \\ + &
    \mu^{-1}_{xy}\left[ E_{0s}t_{ss}^{(i)} k_z'^{(i)} \sin\phi +
      E_{0p}t_{pp}^{(i)} \sqrt{k_{\parallel}^2 + {k_z^{'(i)}}^2}
      \cos\phi + E_{0s}t_{ps}^{(i)} \sqrt{k_{\parallel}^2 +
        {k_z^{'(i)}}^2} \cos\phi + E_{0p}t_{sp}^{(i)}
      k_z'^{(i)}\sin\phi \right] \\ + & \mu^{-1}_{xz} \left[
      E_{0s}t_{ss}^{(i)}k_{\parallel} + E_{0p}
      t_{sp}^{(i)}k_{\parallel} \right] \Bigg\}
   \end{split}
  \end{equation}
  \begin{equation}
  \begin{split}
    \hat{y}: E_{0s} k_z \sin\phi (1 - r_{ss}) + & E_{0p}\cos\phi (1 +
    r_{pp}) + E_{0s}r_{ps} \cos\phi - E_{0p}r_{sp} k_z\sin\phi =
    \\ \sum_{i=1}^{2} \Bigg\{& \mu^{-1}_{yx} \left[ E_{0s}
      t_{ss}^{(i)} k_z'^{(i)} \cos\phi - E_{0p}t_{pp}^{(i)}
      \sqrt{k_{\parallel}^2 + {k_z^{'(i)}}^2} \sin\phi - E_{0s}
      t_{ps}^{(i)} \sqrt{k_{\parallel}^2 + {k_z^{'(i)}}^2} \sin\phi +
      E_{0p} t_{sp}^{(i)} k_z'^{(i)} \cos\phi \right] \\ + &
    \mu^{-1}_{yy}\left[ E_{0s}t_{ss}^{(i)} k_z'^{(i)} \sin\phi +
      E_{0p}t_{pp}^{(i)} \sqrt{k_{\parallel}^2 + {k_z^{'(i)}}^2}
      \cos\phi + E_{0s}t_{ps}^{(i)} \sqrt{k_{\parallel}^2 +
        {k_z^{'(i)}}^2} \cos\phi + E_{0p}t_{sp}^{(i)} k_z'^{(i)}
      \sin\phi \right] \\ + & \mu^{-1}_{yz} \left[ E_{0s}t_{ss}^{(i)}
      k_{\parallel} + E_{0p} t_{sp}^{(i)} k_{\parallel} \right]
    \Bigg\}
  \end{split}
  \end{equation}
\end{subequations}
with $\mu_{ij}^{-1}$ being the elements of the inverse of
$\bar{\bar{\bm \mu}}$ matrix assuming that it is invertible. We get
from the electric field's boundary conditions, taking $E_{0p} = 0$
(incidence of $\bm{s}$-polarized light),
\begin{equation}\label{Eq:Primary_eq_01}
  1 + r_{ss} = \sum_{i=1}^{2}t_{ss}^{(i)} , \quad k_z r_{ps} = -
  \sum_{i=1}^{2} \frac{k_z'^{(i)} t_{ps}^{(i)}}{\sqrt{k_{\parallel}^2
      + {k_z^{'(i)}}^2}} = - \sum_{i=1}^{2} \frac{k_z^{'(i)} A^{(i)}
    t_{ss}^{(i)}}{\sqrt{k_{\parallel}^2 + {k_z^{'(i)}}^2}}
\end{equation}
and, taking $E_{0s} = 0$ (incidence of $\bm{p}$-polarized light),
\begin{equation}\label{Eq:Primary _eq_02}
  k_z (1 - r_{pp}) = \sum_{i=1}^{2}\frac{
    t_{pp}^{(i)}k_z'^{(i)}}{\sqrt{k_{\parallel}^2 + {k_z^{'(i)}}^2}},
  \quad r_{sp} = \sum_{i=1}^{2} t_{sp}^{(i)} = \sum_{i=1}^{2}
  t_{pp}^{(i)} B^{(i)}.
\end{equation}
On the other hand, we get from the boundary conditions of the magnetic
field, setting $E_{0p}=0$ (incidence of $\bm{s}$-polarized light),
\begin{equation}\label{Eq:System_of_Eqs_ss}
  \begin{split}
     \sum_{i=1}^{2} & \left[a_s^{(i)} t_{ss}^{(i)} + 2 k_z \cos\phi
       \right] = 0 \\ \sum_{i=1}^{2} & \left[b_s^{(i)} t_{ss}^{(i)} +
       2 k_z \sin\phi \right] = 0
  \end{split}
\end{equation}
where
\begin{equation}
\begin{split}
   a_s^{(i)} = & -\cos\phi \left[ k_z + \mu_{xx}^{-1}k_z^{'(i)} +
     \sqrt{k_{\parallel}^2 + {k_z^{'(i)}}^2} A^{(i)} \mu_{xy}^{-1}
     \right] \\ - & \sin\phi \left[ -\frac{k_z'^{(i)}}{k_z
       \sqrt{k_{\parallel}^2 + {k_z^{'(i)}}^2} }A^{(i)} +
     k_z'^{(i)}\mu_{xy}^{-1} - \sqrt{k_{\parallel}^2 + {k_z^{'(i)}}^2}
     \mu_{xx}^{-1}A^{(i)} \right] - k_{\parallel} \mu_{xz}^{-1}
 \end{split}
\end{equation}
\begin{equation}
\begin{split}
  b_s^{(i)} = & -\sin\phi \left[ k_z + \sqrt{k_{\parallel}^2 +
      {k_z^{'(i)}}^2} \mu_{xy}^{-1} A^{(i)} + \mu_{yy}^{-1}k_z'^{(i)}
    \right] \\ + & \cos\phi \left[ -\frac{k_z'^{(i)}}{k_z
      \sqrt{k_{\parallel}^2 + {k_z^{'(i)}}^2} }A^{(i)} +
    k_z'^{(i)}\mu_{xy}^{-1} - \sqrt{k_{\parallel}^2 + {k_z^{'(i)}}^2}
    \mu_{yy}^{-1}A^{(i)} \right] - k_{\parallel} \mu_{yz}^{-1}
 \end{split}
\end{equation}
Also, setting $E_{0s} = 0 $ (incidence of $\bm{p}$-polarized light),
we get:
\begin{equation}\label{Eq:System_of_Eqs_pp}
   \begin{split}
     \sum_{i=1}^{2} & \left[a_p^{(i)} t_{pp}^{(i)} - 2 \sin\phi
       \right] = 0 \\ \sum_{i=1}^{2} & \left[b_p^{(i)} t_{pp}^{(i)} +
       2 \cos\phi \right] = 0
  \end{split}
\end{equation}
where
\begin{equation}
\begin{split}
  a_p^{(i)} = & -\sin\phi\left[ -\frac{k_z'^{(i)}}{k_z
      \sqrt{k_{\parallel}^2 + {k_z^{'(i)}}^2} } -
    \sqrt{k_{\parallel}^2 + {k_z^{'(i)}}^2} \mu_{xx}^{-1} +
    \mu_{xy}^{-1} B^{(i)}k_z'^{(i)} \right] \\ - & \cos\phi
  \left[B^{(i)}k_z + B^{(i)} k_z'^{(i)} \mu_{xx}^{-1} +
    \sqrt{k_{\parallel}^2 + {k_z^{'(i)}}^2} \mu_{xy}^{-1}\right] -
  B^{(i)}k_{\parallel} \mu_{xz}^{-1}
\end{split}
\end{equation}
\begin{equation}
\begin{split}
  b_p^{(i)} = & \cos\phi\left[ -\frac{k_z'^{(i)}}{k_z
      \sqrt{k_{\parallel}^2 + {k_z^{'(i)}}^2} } +
    \mu_{xy}^{-1}k_z'^{(i)} B^{(i)} - \sqrt{k_{\parallel}^2 +
      {k_z^{'(i)}}^2} \mu_{yy^{-1}} \right] \\ - & \sin\phi\left[
    B^{(i)}k_z + B^{(i)}k_z'^{(i)}\mu_{yy}^{-1} +
    \sqrt{k_{\parallel}^2 + {k_z^{'(i)}}^2} \mu_{xy}^{-1} \right] -
  B^{(i)}k_{\parallel} \mu_{yz}^{-1}
\end{split}
\end{equation}
Solving these equations by substituting the values of $A^{(i)},
B^{(i)}$, we find the coefficients $t_{ss}^{(i)}$ and $t_{pp}^{(i)}$
from which the required reflection coefficients are obtained. Below,
we look at specific cases that allow further simplification of these
expressions and lead to useful relations between the reflection
coefficients.

\subsection{Gyromagnetic slab with gyrotropy axis perpendicular to surface}
We calculate the coefficients for light incidence in the
$\bm{xz}$-plane such that $\phi=0$ in the expressions obtained in
previous section. This leads to simplified closed-form expressions for
reflection coefficients. For a gyromagnetic slab with gyrotropy axis
along $\bm{z}$ axis, the only nonzero components of the $\mub$ tensor
are $\mu_{xx} = \mu_{yy}$, $\mu_{zz}$ and $\mu_{xy} = -\mu_{yx}$. We
also have $\mu_{xx}^{-1} = \mu_{yy}^{-1} = \frac{\mu_{xx}}{\mu_{xx}^2
  + \mu_{xy}^2}$, $\mu_{xy}^{-1} = \frac{\mu_{xy}}{\mu_{xx}^2 +
  \mu_{xy}^2}$, and $\mu_{zz}^{-1} = \frac{1}{\mu_{zz}}$. Defining
\begin{equation}
\begin{split}
   U^{(i)} = & k_z + \mu_{xx}^{-1} k_z^{(i)'} + \sqrt{k_{\parallel}^2
     + {k_z^{(i)'}}^2} A^{(i)} \mu_{xy}^{-1} \\ V^{(i)} = &
   -\frac{k_z^{(i)'}}{k_z \sqrt{k_{\parallel}^2 + {k_z^{(i)'}}^2}}
   A^{(i)} + k_z^{(i)'} \mu_{xy}^{-1} - \sqrt{k_{\parallel}^2 +
     {k_z^{(i)'}}^2} \mu_{xx}^{-1} A^{(i)}
\end{split}
\end{equation}
and taking $\phi = 0$ (incidence in $\bm{xz}$-plane), we have:
\begin{subequations}
  \begin{equation}
    r_{ps} = - \frac{2}{V^{(2)}U^{(1)} - V^{(1)}U^{(2)}} \left[
      \frac{k_z^{'(1)}A^{(1)}V^{(2)}}{\sqrt{k_{\parallel}^2 +
          {k_z^{'(1)}}^2}} -
      \frac{k_z^{'(2)}A^{(2)}V^{(1)}}{\sqrt{k_{\parallel}^2 +
          {k_z^{'(2)}}^2}} \right]
  \end{equation}
  \begin{equation}
    r_{sp} = \frac{2}{V^{(2)}U^{(1)} - V^{(1)}U^{(2)}} \left[ U^{(2)}
      - U^{(1)} \right]
  \end{equation}
\end{subequations}
From Eqs.~(\ref{Eq:A(i)_01}) and (\ref{Eq:A(i)_02}), it follows after
substituting $\phi = 0 $ that,
\begin{equation}\label{Eq:A(i)_phi0_01}
  A^{(i)} = \frac{\epsilon - \mu_{zz}^{-1}k_{\parallel}^2 -
    \mu_{xx}^{-1} {k_z^{'(i)}}^2}{\mu_{xy}^{-1}
    k_z^{'(i)}\sqrt{k_{\parallel}^2 + {k_z^{'(i)}}^2}} = -
  \frac{k_z^{'(i)}\mu_{xy}^{-1}}{\frac{\epsilon}{\sqrt{k_{\parallel}^2
        + {k_z^{'(i)}}^2}} - \sqrt{k_{\parallel}^2 + {k_z^{'(i)}}^2}
    \mu_{xx}^{-1}}
\end{equation}
From these expressions and using Eqs.~(\ref{Eq:Primary_eq_01}) and
(\ref{Eq:Primary _eq_02}) we find:
\begin{equation}
  r_{sp} = \frac{2}{V^{(2)}U^{(1)} - V^{(1)}U^{(2)}}\frac{k_z^{'(1)} -
    k_z^{'(2)}}{k_z^{'(1)}k_z^{'(2)}} \left( \epsilon - \mu_{zz}^{-1}
  k_{\parallel}^2 \right)
\end{equation}
and
\begin{equation}\label{Eq:rps_closedform_01}
  r_{ps} = \frac{2}{V^{(2)}U^{(1)} - V^{(1)}U^{(2)}}
  \frac{A^{(1)}A^{(2)}}{\sqrt{k_{\parallel}^2+ {k_z^{'(1)}}^2}
    \sqrt{k_{\parallel}^2 + {k_z^{'(2)}}^2}} \epsilon \left(k_z^{'(1)}
  - k_z^{'(2)}\right)
\end{equation}
In the case when the magnetic field is along one of the axes of the
problem, the parameter $d$ in Eq.~(\ref{Eq:d_eigen_eq}) becomes zero
and Eq.~(\ref{Eq:eigen_eq}) becomes a simple second order
equation. Using its solutions, and taking $\phi =0$, we find the
following relations to hold true:
\begin{subequations}
  \begin{equation}
    {k_z^{'(1)}}^2 {k_z^{'(2)}}^2 = (\epsilon - k_{\parallel}^2
    \mu_{zz}^{-1})\left[ \epsilon(\mu_{xy}^2 + \mu_{xx}^2) -
      \mu_{xx}k_{\parallel}^2 \right]
  \end{equation}
  \begin{equation}
    \left( k_{\parallel}^2 + {k_z^{'(1)}}^2
    \right)\left(k_{\parallel}^2 + {k_z^{'(2)}}^2 \right) = \left(
    \epsilon - k_{\parallel}^2\mu_{zz}^{-1} \right)\epsilon\left(
    \mu_{xx}^2 + \mu_{xy}^2 \right) + \mu_{xx}\epsilon k_{\parallel}^2
  \end{equation}
  \begin{equation}
    {k_z^{'(1)}}^2 + {k_z^{'(2)}}^2 = \mu_{xx}\epsilon +
    \mu_{xx}\left( \epsilon - \mu_{zz}^{-1}k_{\parallel}^2 \right) -
    k_{\parallel}^2
  \end{equation}
\end{subequations}
Using these relations and using the first equality of
Eq.~(\ref{Eq:A(i)_phi0_01}), we obtain:
\begin{equation}
  \frac{A^{(1)}A^{(2)}}{\sqrt{k_{\parallel}^2+ {k_z^{'(1)}}^2}
    \sqrt{k_{\parallel}^2 + {k_z^{'(2)}}^2}} = \frac{\epsilon -
    \mu_{zz}^{-1}k_{\parallel}^2}{\epsilon k_z^{'(1)}k_z^{'(2)}}
\end{equation}
Plugging this into Eq.~(\ref{Eq:rps_closedform_01}) we find
\begin{equation}
  r_{ps}(\theta,\phi=0) = r_{sp}(\theta,\phi=0)
\end{equation}
For the case when $\phi = \pi$, all the equations for $\phi = 0 $
remain the same. This means that $r_{sp}(\theta, \phi=0) =
r_{sp}(\theta,\phi=\pi)$. Based on rotational symmetry in $\bm{xy}$
plane, it can be argued that this holds true for all angles
$(\theta,\phi)$ such that:
\begin{equation}
  r_{ps}(\theta,\phi) = r_{ps}(\theta,\phi+\pi) = r_{sp}(\theta,\phi+\pi)
\end{equation}
Furthermore, it can be shown that
\begin{equation}
  t_{ss}^{(2)} = \frac{2k_z V^{(2)}}{V^{(2)}U^{(1)} - V^{(1)}U^{(2)}},
  \quad t_{ss}^{(2)} = - \frac{2k_z V^{(1)}}{V^{(2)}U^{(1)} -
    V^{(1)}U^{(2)}}
\end{equation}
which is valid for $\phi=0$ or $\phi=\pi$. This means that, through
Eqs.~(\ref{Eq:Primary_eq_01}) and (\ref{Eq:Primary _eq_02}),
$r_{ss}(\theta,\phi=0) = r_{ss}(\theta,\phi=\pi)$ and
$r_{pp}(\theta,\phi=0) = r_{pp}(\theta,\phi=\pi)$. Because of
rotational symmetry, the same holds true for other values of $\phi$.
These relations between the reflection coefficients are useful to
derive the second spin-resolved Kirchhoff's law in the main text. 

\subsection{Gyromagnetic slab with the gyrotropy axis parallel to the surface}
We assume the gyrotropy axis to be along $\bm{x}$ axis. In this case
the only nonzero components of the $\mub$ tensor are $\mu_{xx}$,
$\mu_{yy} = \mu_{zz}$, and $\mu_{yz} = -\mu_{zy}$. Considering the
incidence at arbitrary angle $\theta$ but fixed $\phi = 0$, we define:
\begin{equation}
  U_x^{(i)} = k_z + \mu_{xx}^{-1} k_z^{'(i)}, \quad V_y^{(i)} = -
  \frac{k_z^{'(i)} A^{i}}{k_z \sqrt{k_{\parallel}^2 + {k_z^{'(i)}}^2}}
  - \sqrt{k_{\parallel}^2 + {k_z^{'(i)}}^2}A^{(i)}\mu_{xx}^{-1}
\end{equation}
and performing similar analysis as above, we find that:
\begin{equation}
  r_{sp} = \frac{-2}{\left( U_x^{(2)}V_y^{(1)} -U_x^{(1)}V_y^{(2)}
    \right) + k_{\parallel} \mu_{yz}^{-1}\left( U_x^{(1)} - U_x^{(2)}
    \right)} \left( U_x^{(2)} - U_x^{(1)}\right)
\end{equation}
and
\begin{equation}
  r_{ps} = \frac{2}{\left( U_x^{(2)}V_y^{(1)} -U_x^{(1)}V_y^{(2)}
    \right) + k_{\parallel} \mu_{yz}^{-1}\left( U_x^{(1)} - U_x^{(2)}
    \right)} \left[ \frac{k_z^{'(1)}A^{(1)}}{\sqrt{k_{\parallel}^2 +
        {k_z^{'(1)}}^2}}\left(V_y^{(2)} -
    k_{\parallel}\mu_{yz}^{-1}\right) -
    \frac{k_z^{'(2)}A^{(2)}}{\sqrt{k_{\parallel}^2 +
        {k_z^{'(2)}}^2}}\left(V_y^{(1)} -
    k_{\parallel}\mu_{yz}^{-1}\right) \right].
\end{equation}
Simplifying the above expressions, we obtain:
\begin{equation}
  r_{sp} = \frac{2}{\left( U_x^{(2)}V_y^{(1)} -U_x^{(1)}V_y^{(2)}
    \right) + k_{\parallel} \mu_{yz}^{-1}\left( U_x^{(1)} - U_x^{(2)}
    \right)}\mu_{xx}^{-1} \left( k_z^{'(1)} - k_z^{'(2)}\right)
\end{equation}
\begin{equation}
  r_{ps} = -\frac{2}{\left( U_x^{(2)}V_y^{(1)} -U_x^{(1)}V_y^{(2)}
    \right) + k_{\parallel} \mu_{yz}^{-1}\left( U_x^{(1)} - U_x^{(2)}
    \right)}\frac{\epsilon A^{(1)}A^{(2)}}{\sqrt{k_{\parallel}^2 +
      {k_z^{'(1)}}^2}\sqrt{k_{\parallel}^2 + {k_z^{'(2)}}^2} } \left(
  k_z^{'(1)} - k_z^{'(2)} \right)
\end{equation}
Again, we find the following equation holds true:
\begin{equation}\label{Eq:A1A2over12}
  \frac{A^{(1)}A^{(2)}}{\sqrt{k_{\parallel}^2 +
      {k_z^{'(1)}}^2}\sqrt{k_{\parallel}^2 + {k_z^{'(2)}}^2} } =
  \frac{\mu_{xx}^{-1}}{\epsilon}
\end{equation}
This leads to:
\begin{equation}
  r_{ps}(\theta,\phi=0) = -r_{sp}(\theta,\phi=0)
\end{equation}
If $\phi = \pi$, we find that:
\begin{equation}
  r_{sp} = \frac{2}{\left( U_x^{(2)}V_y^{(1)} -U_x^{(1)}V_y^{(2)}
    \right) - k_{\parallel} \mu_{yz}^{-1}\left( U_x^{(1)} - U_x^{(2)}
    \right)} \left( U_x^{(2)} - U_x^{(1)}\right)
\end{equation}
and
\begin{equation}
  r_{ps} = -\frac{2}{\left( U_x^{(2)}V_y^{(1)} -U_x^{(1)}V_y^{(2)}
    \right) - k_{\parallel} \mu_{yz}^{-1}\left( U_x^{(1)} - U_x^{(2)}
    \right)} \left[ \frac{k_z^{'(1)}A^{(1)}}{\sqrt{k_{\parallel}^2 +
        {k_z^{'(1)}}^2}}\left(V_y^{(2)} +
    k_{\parallel}\mu_{yz}^{-1}\right) -
    \frac{k_z^{'(2)}A^{(2)}}{\sqrt{k_{\parallel}^2 +
        {k_z^{'(2)}}^2}}\left(V_y^{(1)} +
    k_{\parallel}\mu_{yz}^{-1}\right) \right].
\end{equation}
following the same procedure discussed above, we obtain:
\begin{equation}
  r_{ps} = \frac{2}{\left( U_x^{(2)}V_y^{(1)} -U_x^{(1)}V_y^{(2)}
    \right) - k_{\parallel} \mu_{yz}^{-1}\left( U_x^{(1)} - U_x^{(2)}
    \right)}\frac{\epsilon A^{(1)}A^{(2)}}{\sqrt{k_{\parallel}^2 +
      {k_z^{'(1)}}^2}\sqrt{k_{\parallel}^2 + {k_z^{'(2)}}^2} } \left(
  k_z^{'(1)} - k_z^{'(2)} \right)
\end{equation}
Eq.~(\ref{Eq:A1A2over12}) still holds for $\phi = \pi$. Therefore, we
get
\begin{equation}
  r_{ps}(\theta,\phi=\pi) = -r_{sp}(\theta,\phi=\pi)
\end{equation}
When $\phi = \pi/2$, the denominator in Eq.~(\ref{Eq:A(i)_01}) and the
numerator in Eq.~(\ref{Eq:A(i)_02}) become zero. In this case,
\begin{equation}
   {k_z^{(1)'}}^2 = \frac{\epsilon}{\mu_{xx}^{-1}} - k_{\parallel}^2,
   \quad {k_z^{(2)'}}^2 = \frac{\epsilon}{\mu_{yy}^{-1}} -
   k_{\parallel}^2
\end{equation}
Using these relations, we find that $t_{ss}^{(1)} = t_{sp}^{(1)} =
t_{pp}^{(2)} = t_{ps}^{(2)}= 0 $. On the other hand, we get from
Eqs.~(\ref{Eq:System_of_Eqs_ss}) and (\ref{Eq:System_of_Eqs_pp}) that
$t_{sp}^{(2)} = t_{ps}^{(1)} = 0$. This means that the only non-zero
transmission coefficients are $t_{ss}^{(2)}$ and $t_{pp}^{(1)}$ and
substituting these coefficients, we get
\begin{equation}
  r_{sp}(\theta,\phi=\pi/2)=r_{ps}(\theta,\phi=\pi/2) = 0
\end{equation}
While we prove the above conditions analytically, for other values of
$\phi$, we have to rely on numerical approach. Interestingly, the
condition $r_{sp}(\theta,\phi)=-r_{ps}(\theta,\phi)$ holds true for
all angles. This condition is used to derive the third spin-resolved
Kirchhoff's law.

While we provide one example here, it is impractical to derive the
closed-form expressions of Fresnel coefficients for arbitrary
incidence angles ($\theta,\phi$) for each bianisotropic material
class. The generic expressions are evidently complicated (see above)
because of the nontrivial dependence of the wavevector and the plane
wave solutions inside each such medium on many parameters and often
through conditional expressions (real or complex roots of quartic
polynomials). Furthermore, such closed form expressions, if obtained,
cannot be readily extended to describe a composite medium such as a
material which is simultaneously gyrotropic and magnetoelectric. The
task of obtaining the closed-form expressions for both reflection and
transmission coefficients for a multilayered planar slab becomes more
complicated because it involves solving the boundary conditions at
many interfaces rather than a single interface for the half-space
geometry. Because of these reasons, we have undertaken a more
practical alternative of using exact numerical methods to obtain the
Fresenel coefficients for arbitrary angles of incidence and for
generic multilayered or composite planar slabs. The various relations
between the reflection and transmission coefficients discovered using
this numerical approach are noted in the manuscript.




\bibliography{photon}